\documentclass[10pt,twocolumn,showpacs,amsmath,amssymb,aps,prx,longbibliography,superscriptaddress,floatfix]{revtex4-1}
\usepackage{times}
\usepackage[utf8]{inputenc}
\usepackage{color}
\usepackage{array}
\usepackage{verbatim}
\usepackage{multirow}
\usepackage{amsmath}
\usepackage{amssymb}
\usepackage{graphicx}
\usepackage{esint}
\usepackage[unicode=true,
 bookmarks=true,bookmarksnumbered=true,bookmarksopen=false,
 breaklinks=true,pdfborder={0 0 0},backref=false,colorlinks=true]
 {hyperref}
\usepackage{enumitem}
\usepackage{cleveref}
\usepackage{bm}
\usepackage{xcolor}
\usepackage[bb=boondox]{mathalfa}
\usepackage[title]{appendix}
\usepackage{makecell}
\usepackage{pifont}
\usepackage{diagbox}

\hypersetup{
    linkcolor=red,          
    citecolor=blue,        
    filecolor=magenta,      
    urlcolor=magenta           
}

\usepackage{cleveref}
\crefname{appendix}{Appendix}{Appendices}
\crefname{equation}{Eq.}{Eqs.}
\crefname{figure}{Fig.}{Figs.}
\crefname{table}{Table}{Tables}
\crefname{section}{Sec.}{Secs.}

\renewcommand{\paragraph}[1]{\vspace{0.2cm}{\bf \textit{#1}}}
\def\ie{{\it i.e.},\ }
\def\eg{{\it e.g.},\ }

\definecolor{Gray}{gray}{0.85}
\newcolumntype{a}{>{\columncolor{Gray}}c}

\usepackage{simpler-wick}
\allowdisplaybreaks[1] 

\newcommand{\mrm}{\mathrm}

\newcommand{\hH}{\hat{H}}

\def\nnorm#1{\left|\left| #1 \right|\right|}

\def\tr{\mathrm{Tr}}

\def\kk{ {\mathbf{k}} }
\def\KK{ {\mathbf{K}} }
\def\bb{\mathbf{b}} 
\def\RR{\mathbf{R}} 
\def\rr{\mathbf{r}} 
\def\dd{\mathbf{d}} 
\def\DD{\mathbf{D}} 
\def\qq{\mathbf{q}}
\def\pp{\mathbf{p}}
\def\QQ{\mathbf{Q}}
\def\oo{\mathbf{0}}
\def\GG{\mathbf{G}}
\def\ttau{\bm{\tau}}

\def\MBZ{\mrm{MBZ}}
\def\fMBZ{\mrm{fMBZ}}
\def\mS{\mathcal{S}}
\begin{document}
\title{Correlated insulators and charge density wave states in chirally twisted triple bilayer graphene}

\author{Geng-Dong Zhou}
\affiliation{International Center for Quantum Materials, School of Physics, Peking University, Beijing 100871, China}

\author{Yi-Jie Wang}
\affiliation{International Center for Quantum Materials, School of Physics, Peking University, Beijing 100871, China}

\author{Wen-Xuan Wang}
\affiliation{International Center for Quantum Materials, School of Physics, Peking University, Beijing 100871, China}

\author{Xiao-Bo Lu}
\affiliation{International Center for Quantum Materials, School of Physics, Peking University, Beijing 100871, China}
\affiliation{Collaborative Innovation Center of Quantum Matter, Beijing 100871, China}

\author{Zhi-Da Song}
\email{songzd@pku.edu.cn}
\affiliation{International Center for Quantum Materials, School of Physics, Peking University, Beijing 100871, China}
\affiliation{Collaborative Innovation Center of Quantum Matter, Beijing 100871, China}
\affiliation{Hefei National Laboratory, Hefei 230088, China}

\date{\today}
\begin{abstract}
Motivated by recent experimental observations of displacement-field-tuned correlated insulators at integer and half-integer fillings in chirally twisted triple bilayer graphene (CTTBG), we study the single-particle and interacting physics of CTTBG. We find that there are two inequivalent stacking orders, {\it i.e.}, ABABBC and ABABAB, and both exhibit flat bands with nontrivial topology. We then use the Hartree-Fock approximation to calculate the rich phase diagram of CTTBG at all integer and half-integer fillings in both stacking orders and under the vertical displacement field. Under a small displacement field, the groundstates are flavor polarized states for ABABBC stacking order and intervalley coherent states for ABABAB stacking order at all integer and half-integer fillings. A larger displacement field will turn them into layer-polarized states. At half-integer fillings, the groundstates also exhibit charge density wave (CDW) order. For ABABAB stacking, the groundstates are always $2\times1$ stripe state among a range of displacement fields. For ABABBC stacking, the groundstates are also $2\times1$ stripe states under a small displacement field and a larger displacement will possibly favor further translation-symmetry-breaking, depending on filling and the direction of the displacement field. We demonstrate that the CDW states observed in the experiment can originate from the strong Coulomb interaction of the flat band electrons.
\end{abstract}
\maketitle

\section{Introduction}
Moir\'e materials have drawn lots of attention as they provide highly tunable platforms to demonstrate the interplay of strong correlation effect and topology, which lead to exotic phenomena including correlated insulators \cite{cao_unconventional_2018,lu_superconductors_2019,yankowitz_tuning_2019,sharpe_emergent_2019,wong_cascade_2020,stepanov_untying_2020}, unconventional superconductivity \cite{cao_unconventional_2018,lu_superconductors_2019,yankowitz_tuning_2019,stepanov_untying_2020,saito_independent_2020,oh_evidence_2021}, Chern insulators \cite{das_symmetry-broken_2021,serlin_intrinsic_2020,grover_chern_2022}, fractional Chern insulators \cite{cai_signatures_2023,park_observation_2023,xu_observation_2023,zeng_thermodynamic_2023}, \textit{etc}.
A natural generalization of the first Moir\'e materiel magic-angle twisted bilayer graphene (TBG) is the family of twisted multilayer graphene \cite{khalaf_magic_2019,liang_moire_2022}, including alternatingly and chirally twisted trilayer graphene \cite{park_tunable_2021,cao_pauli-limit_2021,hao_electric_2021,park_robust_2022,kim_evidence_2022,turkel_orderly_2022,liu_isospin_2022,shen_dirac_2023}, twisted double bilayer graphene \cite{koshino_band_2019,shen_correlated_2020,liu_tunable_2020,wu_ferromagnetism_2020,he_symmetry_2021,he_symmetry-broken_2023,su_superconductivity_2023}, \textit{etc.}, where superconductivity, correlated insulators, \textit{etc.}, have also been observed.
For multilayer Moir\'e material with more than one twisting interface, there will be multiple Moir\'e patterns and the stacking order of these Moir\'e patterns will play a significant role, which has been found in chirally twisted trilayer graphene \cite{devakul_magic-angle_2023,kwan_strong-coupling_2023}. Among these multilayer graphene families, there is another interesting twisted multilayer Moir\'e material called chirally twisted triple bilayer graphene (CTTBG), which is formed by chirally stacked three Bernal bilayer graphenes with the same twist angle. It has been suggested to own robust flat bands \cite{liang_moire_2022} and studied in the recent experiments \cite{wang_correlated_2024}, where correlated gapped states at integer and half-integer fillings tuned by displacement field and magnetic field are found. Notably, Ref.\cite{wang_correlated_2024} suggests that the correlated gapped states at non-integer fillings are charge density wave (CDW) states. A thorough study of the single-particle and interacting physics of CTTBG is needed to understand these insulating states observed in experiments.

In this work, we perform systematic single particle and Hartree-Fock study on CTTBG at all integer and half-integer fillings at different displacement fields for two inequivalent stacking orders, which are denoted as ABABAB and ABABBC as detailed in \cref{sec:model}. 
At the single-particle level, we find that CTTBG in both stacking at $\theta=1.70^\circ$ possess a pair of flat bands separated from other bands, exhibiting non-trivial topology. In ABABBC-stacking CTTBG, we predict that one can also observe the valley hall effect when the Fermi level lies in the band gaps at fillings $\pm 4$. 
We also find that the flat band electrons in CTTBG are distributed dispersedly, in contrast to TBG where the flat band electrons are highly localized at the AA regions \cite{song_magic-angle_2022,shi_heavy-fermion_2022,zhou_kondo_2024,wang_molecular_2024}. This underpins the significance of the nonlocal nature of Coulomb interactions and is consistent with the observed charge density wave states. 
Using the Hartree-Fock approximation, the groundstates at zero displacement field are found to be intervalley coherent (IVC) states in ABABAB stacking and flavor polarized states in ABABBC stacking. At a large displacement field, layer-polarized states are favored. At half-integer fillings, we find that groundstates further break the translation symmetry of the Moir'e superlattice and this translation-symmetry-breaking can be tuned by the displacement field.

\section{Model and method\label{sec:model}}
In this section, we discuss the stacking order of CTTBG and briefly summarize the model and notation we used.
\subsection{Geometry and stacking order}
We consider CTTBG consisting of three Bernal-stacking bilayer graphenes twisted by angle $(\theta,0,-\theta)$ from top to bottom, and the experimental twist angle $\theta$ is $1.70^\circ$ \cite{wang_correlated_2024}. The unrotated lattice vectors of the bilayer graphene are $a_0(\pm \frac{1}{2},\frac{\sqrt{3}}{2})$ where $a_0=2.46${\AA} is the lattice constant. CTTBG possesses two Moir\'e superlattices with lattice constant $a_M = a_0/(2\sin\frac{\theta}{2})$, one formed by the top and middle layers and one by the middle and bottom layers. At the Moir\'e scale, the stacking order of CTTBG is decided by the displacement $\mathbf{D}$ of the bottom Moir\'e pattern relative to the top Moir\'e pattern. 
These two Moir\'e patterns are also twisted by $\theta$, leading to a moir\'e-of-moir\'e structure with lattice constant $\sim a_0/\theta^2\approx 280 \mrm{nm}$ when $\theta = 1.70^\circ$ and the displacement $\DD$ will be position-dependent. However, similar to chirally and alternatingly twisted trilayer graphene \cite{nakatsuji_multi-scale_2023,kwan_moire_2023,kwan_strong-coupling_2023}, one expects that lattice relaxation will reconstruct the moir\'e-of-moir\'e lattice into domains which contain only single Moir\'e structure and they are separated by domain walls which form triangular or hexagonal networks. Within one domain, $\DD$ is uniform and takes a fixed value that is favored by lattice relaxation. 

In this work, we consider a single domain with uniform displacement $\mathbf{D}=(0,0),\pm a_M (\frac{1}{\sqrt{3}},0) $ modulo the Moir\'e lattice vector, where CTTBG respects $C_{3z}$ symmetry and the elastic energy reaches the local extremum. These three cases correspond to the ABAB-stacking region in the top Moir\'e pattern aligning with the ABAB/ABBC/ABCA-stacking region in the bottom Moir\'e pattern, respectively, and we denote them as ABABAB, ABABBC and ABABCA stacking orders. 
Besides, $C_{2x}$ is preserved by each Bernal bilayer and swaps the top and bottom Moir\'e patterns, thus changing CTTBG with displacement $\DD$ to $-C_{2x}\DD$. Therefore, the ABABAB stacking order further preserves $C_{2x}$ symmetry, and the later two stacking orders can be transformed to each other under $C_{2x}$, for which we only need to consider two inequivalent stacking orders ABABAB and ABABBC, as shown in \cref{fig:singleparticle}(a). There is no other spatial symmetry, hence the space groups are just $D_3$ and $C_{3z}$ for ABABAB-stacking and ABABBC-stacking CTTBG, respectively.

The favored displacement $\mathbf{D}$ should be determined by minimizing the total energy. In Ref.\cite{nakatsuji_multi-scale_2023}, the authors pointed out that in twisted trilayer systems, whether a stacking order is favored can be inferred from whether the top and the bottom layer introduce the same or opposite tendency of deformation on the middle layer. As an example, we review the twisted trilayer graphene (TTG) cases in Ref.\cite{nakatsuji_multi-scale_2023} here. We denote the local twist angles as $(\theta_1,\theta_2,\theta_3)$ for the three layers, which are position-dependent as deformation is allowed.
For each pair of adjacent layers, since AA regions cost more energy, the local twist angle tends to increase in AA regions and decrease in AB/BA regions, in order that AA regions shrink and AB/BA regions expand to save energy. Thus, for chiral TTG ($\theta_1>\theta_2>\theta_3)$, if AA region aligns AA region lattice relaxation tends to increase both $\theta_1-\theta_2$ and $\theta_2-\theta_3$, which make $\theta_2$ in dilemma. By contrast, if the AA region aligns with the AB region, one can simultaneously increase $\theta_1-\theta_2$ and decrease $\theta_2-\theta_3$. As a result, AAB stacking is preferred over AAA stacking in chiral TTG. Conversely, in alternating TTG ($\theta_1>\theta_2,\theta_3>\theta_2)$ AAA stacking is preferred. These are confirmed by numerical calculation in \cite{nakatsuji_multi-scale_2023,devakul_magic-angle_2023}. 
Based on the same principle, we do not perform lattice relaxation calculation but just argue that it also prefers ABABBC-stacking order in CTTBG. We list the stacking region and the tendency of deformation in \cref{tab:deformation}, which shows that the deformations of two Moir\'e patterns are frustrated in all three stacking regions in ABABAB-stacking CTTBG and in only one stacking region in ABABBC-stacking CTTBG, akin to chiral TTG. However, this is not a quantitative calculation, and the corrugation, the energy of the electron, and the electron-phonon coupling are not fully considered, so the ABABAB-stacking order is not ruled out. Thus, we performed calculations on both stacking orders in this work.

\begin{table}[h]
\begin{tabular}{c|ccccc|ccccc}\hline\hline
 & \multicolumn{4}{c|}{ABABAB} & Favorable  & \multicolumn{4}{c|}{ABABBC} & Favorable\\ \hline
twist angle & \multicolumn{1}{c|}{\;\;$\theta$\;\;}   & \multicolumn{2}{c|}{\;$\delta\theta$\;}       &   \multicolumn{1}{c|}{\;$-\theta$\;} & \diagbox{}{}   & \multicolumn{1}{c|}{\;\;$\theta$\;\;}   & \multicolumn{2}{c|}{\;$\delta\theta$\;}   &                  \multicolumn{1}{c|}{\;$-\theta$\;}   & \diagbox{}{}\\ \hline
 \makecell{relative\\ twist angle} & \multicolumn{2}{c|}{$\theta-\delta\theta$}              & \multicolumn{2}{c|}{$\theta+\delta\theta$} & \diagbox{}{} &\multicolumn{2}{c|}{$\theta-\delta\theta$}              & \multicolumn{2}{c|}{$\theta+\delta\theta$} & \diagbox{}{}\\ \hline
 Region & \multicolumn{2}{c|}{AA$(\uparrow)$}              & \multicolumn{2}{c|}{AA$(\uparrow)$} & $\times$ & \multicolumn{2}{c|}{AA$(\uparrow)$}              & \multicolumn{2}{c|}{AB$(\downarrow)$}    &\checkmark          \\ \cline{2-11}
 & \multicolumn{2}{c|}{BA$(\downarrow)$}              & \multicolumn{2}{c|}{BA$(\downarrow)$} & $\times$ & \multicolumn{2}{c|}{BA$(\downarrow)$}              & \multicolumn{2}{c|}{AA$(\uparrow)$}          &\checkmark    \\ \cline{2-11}
 & \multicolumn{2}{c|}{AB$(\downarrow)$}              & \multicolumn{2}{c|}{AB$(\downarrow)$} &$\times$ & \multicolumn{2}{c|}{AB$(\downarrow)$}              & \multicolumn{2}{c|}{BA$(\downarrow)$}        &$\times$      \\ \hline\hline
\end{tabular}
\caption{The local deformation of each layer. The second row shows the twist angle of each bilayer, and the third row shows the twist angle of the top layer relative to the bottom layer in each Moir\'e pattern. The 4-6th rows show the stacking regions and the tendency of deformation of the local relative twist angles. The stacking regions are labeled only by the atoms in adjacent monolayers instead of bilayers, \eg by BA-stacking instead of ABAB-stacking. $\uparrow$ ($\downarrow$) means that the local relative twist angle tends to increase (decrease).  \label{tab:deformation}}
\end{table}

\subsection{Non-interacting model\label{sec:non-int-model}}
\begin{figure}
    \centering
    \includegraphics[width=\linewidth]{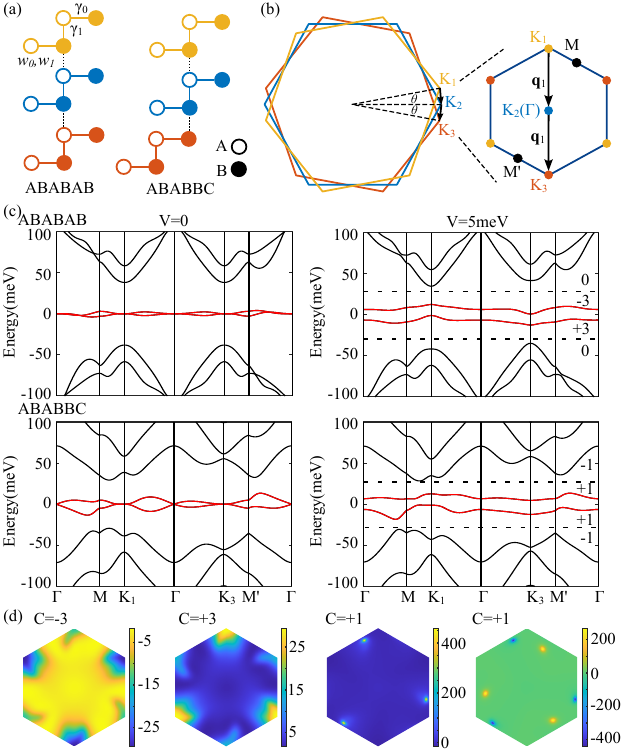}
    \caption{(a) Geometry of the two stacking orders, viewed from x direction. The couplings are also indicated here. (b) Left: The graphene Brillouin zone. Right: the MBZ. (c) The non-interacting bands in $\eta=+$ valley. The stacking orders in the upper and lower panels are ABABAB and ABABBC, respectively, and the displacement fields applied are $V=0$ and $V=5$meV for the left and right panels. The numbers mark the total Chern number of the remote conductance bands, the flat conductance band, the flat valence band, and the remote valence bands from top to bottom, respectively. (d) From left to right: the Berry curvature of conductance, valence flat band in ABABAB stacking and conductance, valence flat band in ABABBC stacking under $V_d=5$meV. The Berry curvature is normalized in the convention that the area of MBZ is $4\pi^2$.}
    \label{fig:singleparticle}
\end{figure}

Here we briefly summarize the non-interacting model and more details can be found in \cref{app:model}. 
With the $i$-th bilayer (counted from top to bottom) displaced by $\mathbf{d}_i$, the non-interacting continuum model \cite{khalaf_magic_2019,bistritzer_moire_2011,lee_theory_2019} of CTTBG reads
\begin{align}
    \hat{H}_0 &= \hat{H}_{BM} + \hat{H}_V \, ,\\
    \hat{H}_{BM}&=\int{\rm d}^2\rr \Bigg[\sum_{i=1}^{3}\sum_{\eta s,aa'}c^\dagger_{i\rr a\eta s}h^{\eta}_{aa'}(-i\mathbf{\nabla}_{\rr})c_{i\rr a'\eta s}+\nonumber\\
    &\quad \left(\sum_{i=1}^{2}\sum_{\eta,s}\sum_{aa'}[T^{i,i+1}_{\eta}(\rr)]_{aa'}c^\dagger_{i\rr a\eta s}c_{i\rr a'\eta s}+h.c.\right)\Bigg]\, ,\label{eq:HBM-realspace}
\end{align}
where 
\begin{align}
    h^{\eta}(\kk) &= \frac{\sqrt{3}}{2}\gamma_0a_0\sigma_0\otimes(\eta k_x \sigma_x + k_y\sigma_y) + \left(\begin{array}{cccc}
        0 & 0 & 0 &\gamma_1\\
       0  & 0 & 0 & 0\\
       0 & 0 & 0 & 0 \\
       \gamma_1 & 0 &0 & 0
    \end{array}\right)
\end{align}

 is the low energy Hamiltonian of Bernal-stacking bilayer graphene near the Dirac cone at $\eta\mathbf{K}=\eta (\frac{4\pi}{3a_0},0),\eta=\pm$, $\sigma_{0,x,y}$ are the Pauli matrices and 
\begin{align}
    [T^{i,i+1}_{\eta}(\rr)]_{aa'}= \sum_{n}\left[\left(\begin{array}{cc}
    0 & 0 \\
    1 & 0 
    \end{array}\right)\otimes T_n\right]_{aa'}e^{i\eta\qq_n\cdot[\rr-\DD_{i,i+1}]}
\end{align}
describes the Moir\'e hopping. We adopt the hopping parameters $\gamma_0=2610\mrm{meV},\gamma_1=361\mrm{meV}$, which are defined in \cref{fig:singleparticle}(a). $T_n=w_0\sigma_0+w_1\sigma_x\cos\frac{2(n-1)\pi}{3}+w_1\sigma_y\sin\frac{2(n-1)\pi}{3}$ where $w_0=88\mrm{meV},w_1=110\mrm{meV}$ are AA and AB/BA hopping, respectively. The sublattice index $a(a')$ labels the atoms in a unit cell with order (1A,1B,2A,2B). $c_{i\rr a\eta s}$ annihilates an electron in the $i$-th bilayer at position $\rr$ in $\eta$ valley with spin $s$ and sublattice index $a$. $\qq_n=C_{3z}^{n-1}k_\theta(0,-1)$, where $k_\theta=2|\KK|\sin\frac{\theta}{2}$.
$\mathbf{D}_{i,i+1}=\frac{\hat{z}}{2\sin(\theta/2)} \times (\bm{\delta}_1+\dd_i-\dd_{i+1})$ is the location of the AA region in the Moir\'e pattern formed by the $i$-th and $(i+1)$-th bilayer, $\bm{\delta}_1=-\frac{a_0}{\sqrt{3}}(0,1)$ is the position of A sublattice relative to the B sublattice in the same unit cell and the same layer, and the displacement of two Moir\'e patterns is $\DD = \DD_{23}-\DD_{12}$. We also neglect other terms like long-range hopping and leave the discussion of their influence to \cref{app:model}. 

Additionally, we add a layer-dependent on-site potential $\hH_V$ which brings $V_d(l-\frac{7}{2})$ for the $l$-th monolayer graphene, which reflects both the gate-induced displacement field and internal displacement field from charge redistribution among layer. In principle, the potential brought by the latter one is not linear with $l$ \cite{ghazaryan_multilayer_2023,avetisyan_electric_2009} but we neglect this effect for simplicity \cite{dong_anomalous_2023}. The space groups of both stacking orders are $C_{3z}$ with displacement field since it does not respect $C_{2x}$ symmetry.

We also summarize the symmetry of the model, as shown in \cref{tab:sym}.
Despite the space group symmetry mentioned above, the model also respects U(2)$\times$U(2) symmetry, consisting of charge U(1)$_c$, valley U(1)$_V$, and spin SU(2)$_{\pm}$ in $\eta=\pm$ valley, as we have neglected spin-orbital couplings and Umklamp scattering term as done in previous work \cite{bultinck_ground_2020,zhang_correlated_2020,liu_nematic_2021,kwan_kekule_2021,christos_correlated_2022,wagner_global_2022,xie_phase_2023}. 

Furthermore, when $\DD_{i,i+1}$ lies in $x$ direction, \ie $\dd_i-\dd_{i+1}$ lies in $y$ direction, which is satisfied for both ABABAB and ABABBC stacking orders, \cref{eq:HBM-realspace} also has an approximate particle-hole (PH) symmetry $\mathcal{P}_y$ which anticommutes with the single particle Hamiltonian $H_{BM}$. Similar to the PH symmetry in TBG \cite{song_all_2019}, $\mathcal{P}_y$ will be broken when quadratic terms of $\kk$ or $\theta$-dependence are included in $h^\eta(\kk)$. $\mathcal{P}_y$ is antiunitary and its matrix representation is $D_{ia\eta,i'a'\eta'}(\mathcal{P}_y)=(-1)^{l_ia}\delta_{aa'}\delta_{\eta\eta'}$ where $l_{ia}=1,\ldots 6$ labels the monolayer, to which the $a$ sublattice in the $i$-th bilayer belongs, from top to bottom. $\mathcal{P}_y$ also changes $\rr$ to $M_{y}\rr$, and hence $\kk$ to $C_{2y}\kk$. One can directly verify the anticommuting relation between $\mathcal{P}_y$ and the intra-bilayer single-particle Hamiltonian, and also the Moir\'e coupling terms by making use of $(T_1)^*=T_1,(T_2)^*=T_3, M_y \qq_1=-\qq_1, M_y\qq_2 = -\qq_3$ and $\qq_1\cdot \DD_{i,i+1}=0,\qq_2\cdot \DD_{i,i+1} = -\qq_3\cdot \DD_{i,i+1}$ given $\DD_{i,i+1}$ lies in $x$ direction.
Besides, $\mathcal{P}_y$ commutes with the single particle Hamiltonian $H_{V}$, so at finite displacement field, the model is not invariant under $\mathcal{P}_y$ but is related to the case with the displacement field reversed, \ie CTTBG at filling $\nu$ under displacement field $V_d$ is equivalent to CTTBG at filling $-\nu$ under displacement field $-V_d$.

The model also respects the spinless time-reversal symmetry (TRS) $T_0=\tau_x K$, which $T_0^2=1$. Combined with U(1)$_V$ we can define another TRS-like symmetry $T_K=e^{-i\pi\tau_z/2}T_0=\tau_y K$ where $T_K^2=-1$.
\begin{table}[h]
    \centering
    \begin{tabular}{c|c|c|c|c}\hline\hline
        Stacking & $V_d$ & Space Group & $\mathcal{P}_y$ & On-site  \\\hline
         \multirow{2}*{ABABAB} & 0 & $D_3$ & \checkmark &\multirow{4}*{\makecell{U(2)$\times$U(2),\\TRS}}\\
         \cline{2-4}& $\neq 0$  & $C_{3z}$ & $\times$ \\
         \cline{1-4}\multirow{2}*{ABABBC} & 0 & $C_{3z}$ & \checkmark \\
         \cline{2-4}& $\neq 0$ & $C_{3z}$ & $\times$ \\\hline\hline
    \end{tabular}
    \caption{Symmetry of the model}
    \label{tab:sym}
\end{table}

\subsection{Interacting model and Hartree-Fock method\label{sec:HF}}
In this subsection, we discuss the interacting model and the Hartree-Fock (HF) method. The total Hamiltonian is $\hat{H}=\hH_{0}+\hH_I$ where  $\hH_I=\frac{1}{2\Omega_{tot}}\sum_{\qq\in \MBZ}\sum_{\GG}V(\qq+\GG)\delta\rho_{\qq+\GG}\delta\rho_{-\qq-\GG}$
where $\delta\rho_{\qq+\GG}, V(\qq+\GG)$ is the $\qq+\GG$ component in the momentum space of the density operator relative to the charge background and the electron-electron interaction, $\Omega_{tot}$ is the total area of the sample and MBZ is the Moir\'e Brillouin zone, as shown in \cref{fig:singleparticle}(b). We adopt the double-gate-screened Coulomb interaction $V(\qq)=\frac{e^2}{2\epsilon|\qq|}\tanh{\frac{\xi |\qq|}{2}}$, where $\xi=30$nm is the distance between the top and the bottom gates, $\epsilon\sim 6$ is the dielectric constant of hBN.

To determine the groundstates at various fillings, we perform a projected HF calculation on the flat bands similar to previous works \cite{bultinck_ground_2020,zhang_correlated_2020,liu_nematic_2021,kwan_kekule_2021,christos_correlated_2022,wagner_global_2022,xie_phase_2023}, whose details can be found in \cref{app:proj-HF}. Diagonalizing $\hH_{0}$ we obtain a set of eigeneneregy $\epsilon_{\kk m\eta}$ and eigenstate $c_{\kk m\eta s}$ where $m$ is the band index. The total Hamiltonian in this basis is 
{\small
\begin{align}
    \hH &= \sum_{\kk n \eta s}\epsilon_{\kk m\eta} c^\dagger_{\kk m\eta s} c_{\kk m\eta s} + \frac{1}{2\Omega_M}\sum_{\kk\kk'\qq}\sum_{\substack{mm'nn'\\\eta\eta' ss'}}\nonumber \\
& U^{\eta\eta'}_{mn,m'n'}(\qq;\kk
,\kk') :c^\dagger_{\kk+\qq m\eta s}c_{\kk n\eta s}::c^\dagger_{\kk'-\qq m'\eta' s'}c_{\kk' n'\eta' s'}: \label{eq:Hproj}
\end{align}
}
where {\small $:c^\dagger_{\kk m\eta s}c_{\kk' m'\eta' s'}: $ is defined as $c^\dagger_{\kk m\eta s}c_{\kk' m'\eta' s'}-\frac{1}{2}\delta_{\kk\kk'}\delta_{mm'}\delta_{\eta\eta'}\delta_{ss'}$} to avoid double counting. Since the parameters we used are from DFT calculation which already contains the self-consistent Hartree potential, we use the high-temperature subtraction scheme where the interaction in the density is measured with respect to the reference density with all bands half-filling at infinite temperature. In this scheme projecting \cref{eq:Hproj} to the flat bands can be simply achieved by restricting $m$ to the flat bands, as detailed in \cref{app:proj-HF}, and we label them as $m=1,2$ hereafter.

To discuss possible translation-symmetry-breaking order, we enlarge the unit cell, which yields a folded Moir\'e Brillouin zone (fMBZ). Each type of translation-symmetry-breaking order is labeled by the fMBZ and its reciprocal lattice vectors $\QQ_1,\QQ_2$, where electrons can be scattered into states with momenta differed by an integer multiple of $\QQ_{1,2}$. We rewrite the momentum in MBZ as $\kk+\QQ_b$ where $\kk$ lies in fMBZ and $\QQ_b$ lies in MBZ and is an integer multiple of $\QQ_{1,2}$. We list the fMBZs we consider in \cref{fig:CDW}(a).
We then perform stanstard HF decomposition for \cref{eq:Hproj}, where the order parameters are $O_{bm\eta s,b'm'\eta's'}=\langle c^\dagger_{\kk+\QQ_b m\eta s}c_{\kk+\QQ'_b m'\eta' s'}\rangle-\frac{1}{2}\delta_{bb'}\delta_{mm'}\delta_{\eta\eta'}\delta_{ss'}$. We start from random order parameters ($\sim$100 samples) and self-consistently calculate order parameters and total energies until they converge. Furthermore, we use the optimal damping method introduced by \cite{cances_can_2000} to accelerate the convergence.

Here, we introduce several quantities to describe the symmetry of the mean-field solution. We first choose the gauge such that the spin up and the spin down wavefunction are the same and the wavefunction in $\eta=-$ valley are related to the wavefunction in $\eta=+$ valley by TRS, as detailed in \cref{app:proj-HF}. There is a remaining gauge degree of freedom that one can choose the basis to be any linear combination of the two flat bands. We work either in the non-interacting energy band basis or the Chern basis introduced later, and the symmetry-breaking quantities we defined here do not depend on this gauge choice. Unless otherwise specified, we use $\sigma_\mu,\tau_\mu,\varsigma_\mu,\mu=0,x,y,z$ to refer to Pauli matrices acting on band index, valley, and spin degree of freedom.
For continuous symmetry operation $e^{-i\theta \hat\Theta}$ with generator  \begin{align}\hat\Theta = \frac{1}{2}\sum_{\kk\in\fMBZ}\sum_{b}\sum_{\substack{m\eta s\\ m'\eta's'}}\Theta_{m'\eta's',m\eta s}c^{\dagger}_{k+\QQ_b m \eta s}c_{k+\QQ_b m' \eta' s'} 
\end{align}
where $\Theta$ will be defined for each symmetry later, the symmetry-breaking strength can be defined by the norm of the commutator of the order parameter and the generator

\begin{align}
    \mS_{\Theta}=\frac{1}{N_M}\sum_{b}\sum_{\kk}\left(||[O_{bb}(\kk),\Theta]||\right)^2\label{eq:ssb-cont}
\end{align}

where $||\cdot||$ is defined by $||A||\equiv\sqrt{\tr(A^\dagger A)}$ for any matrix $A$, and $O_{b''b'}(\kk)$ in \cref{eq:ssb-cont} is an eight-by-eight matrix defined by $\left(O_{b''b'}(\kk)\right)_{m\eta s,m'\eta's'}=O_{b''m\eta s,b'm'\eta's'}(\kk)$ (only $b'=b''=b$ used here).  For valley U(1) $\Theta=\sigma_0\tau_z \varsigma_0$ and we denote the breaking strength as $\mS_V$; We define $\mS_{s\pm} = \frac{1}{4}\sum_{i=x,y,z} \mS_{\sigma_0\tau_\pm \varsigma_{i}}$ for SU(2) within $\pm$ valley, where 
$\sigma_i,\tau_i,\varsigma_i$ are the Pauli matrices act on the spin/valley degree of freedom, and $\tau_{\pm}=\frac{\tau_0\pm\tau_z}{2}$. For example, if the order parameter takes the form $\frac{1}{2}\tau_x\sigma_0\varsigma_0$ at each $\kk$, which means that $\nu=0$ and there are 4 electrons per unit cell among flat bands occupying the equal superposition of two valleys, one will have $\mathcal{S}_V=8,\;\mathcal{S}_{s\pm}=0$; if the order parameter is $\frac{\sigma_0+\sigma_z}{2}\frac{\tau_0+\tau_z}{2}\frac{\varsigma_0+\varsigma_z}{2}-\frac{1}{2}\sigma_0\tau_0\varsigma_0$, which means there is one spin-up electron occupying $\eta=+$ valley, $\mS_V=\mS_{s-}=0$ and $\mS_{s+}=1$. Furthermore $\mS_V,\mS_{s\pm}$ are invariant under all of charge/valley U(1) and spin SU(2)$_\pm$ transformation and independent of the gauge choice among two flat bands, which are proved in \cref{app:order-sym-prop}. 
\section{Results}
\subsection{Single particle band, topology and valley Hall effect\label{sec:singleparticle}}
\begin{figure}
    \centering
    \includegraphics[width=\linewidth]{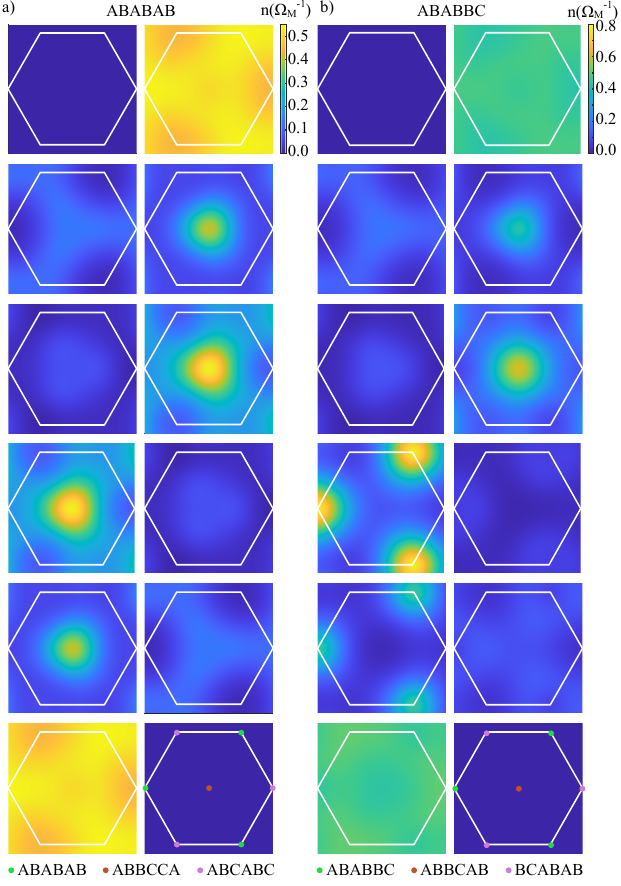}
    \caption{The density profile of filled flat bands for (a) ABABAB stacking and (b) ABABBC stacking order. The density distribution on the A/B sublattice and each monolayer graphene from top to bottom also lie from left to right and from top to bottom, respectively. The regions of different stacking orders are marked in the last figure of each subfigure. Notice that the color codes are the same for the same stacking order and different for different stacking orders, and the origin point adapted in the model in \cref{sec:non-int-model,sec:HF} is located in the ABABAB or ABABBC region.}
    \label{fig:density}
\end{figure}
We plot the non-interacting bands in $\eta=+$ valley for the two stacking orders with and without vertical field in \cref{fig:singleparticle}(c). We find that all of them have two flat bands near the Fermi level separated from other bands, and in ABABAB-stacking CCTBG, the bands are flatter. The bandwidth is around 2meV for the ABABAB-stacking order and around 10meV for the ABABBC-stacking order. 
The total Chern numbers of the two flat bands in $\eta=+$ valley are 0 and 2 for ABABAB and ABABBC-stacking CTTBG, respectively. 
The total Chern numbers for the remote valence bands in $\eta=+$ valley are $0$ and $-1$ for ABABAB and ABABBC-stacking CTTBG, respectively. Thus, for ABABBC-stacking CCTBG at fillings $\nu=\pm 4$, the valley Chern number for filled bands is $\pm 1$, implying the valley Hall effect. 
We also plot the real space density distribution of the filled flat bands in \cref{fig:density}. We find that the density is uniform in the top and bottom layers, and is more concentrated in the AA stacking region of the Moir\'e pattern formed by the twisted adjacent layers in the middle layers. The AA stacking region of the top Moir\'e pattern is aligned with the AA stacking region of the bottom Moir\'e pattern in ABABAB-stacking CTTBG, and the AB stacking region in ABABBC-stacking CTTBG, and so are the localized centers of charge density distribution, as shown by \cref{fig:density}.

Applying a vertical electric field, the two flat bands are separated, and each of them may have nontrivial topological properties. The Chern numbers of valence/conductance flat band are $+3/-3$  and $+1/+1$ in these two stackings, respectively, if we apply an electric field in $z$ direction, as shown in \cref{fig:singleparticle}(c) and $-3/+3$ and $+1/+1$ with the electric field at $-z$ direction. The $\pm 3$ Chern band in ABABAB-stacking CTTBG has already been reported in Ref.\cite{liang_moire_2022}. We also plot the berry curvature of each band in \cref{fig:singleparticle}(d)(e), where one can see that the berry curvature is more uniform in ABABAB-stacking and more concentrated in ABABBC-stacking, which means that if both stackings are experimentally achievable the previous may act as a better candidate to realize fractional Chern insulator \cite{parameswaran_fractional_2012,roy_band_2014}. 
Here we emphasize that as the gap between the two flat bands is small, effects like next nearest neighbor hoppings may close this gap and let the Chern number for each band be ill-defined, which is shown in \cref{fig:singleparticle-nnn}(b) in \cref{app:model}. In either stacking, the gaps between the flat bands and the remote bands are larger; the valley Hall effects in ABABBC stacking at $\nu=\pm 4$ are more robust, as one can see in \cref{fig:singleparticle-nnn}(b) that after adding other terms, the total Chern number of the flat bands and the remote conduction/valence bands are invariant separately.

We also diagonalize the sublattice operators in the flat band subspace in the absence of displacement field and label them by $(\tau,\sigma,s)$, where $\sigma=$A(B) corresponds to the bands with more components from A(B) sublattice, yielding a Chern band basis similar to those used in \cite{bultinck_ground_2020,song_twisted_2021,bernevig_twisted_2021,ledwith_strong_2021,kwan_strong-coupling_2023}. The Chern number of these bands are: $C_{\pm,A,s}=\mp3,C_{\pm,B,s}=\pm 3$ for ABABAB stacking order and $C_{\pm,A,s}=\mp2,C_{\pm,B,s}=\pm 4$ for ABABBC stacking order.

\subsection{Correlated insulator states at integer fillings}

\begin{figure*}
    \centering
    \includegraphics[width=\linewidth]{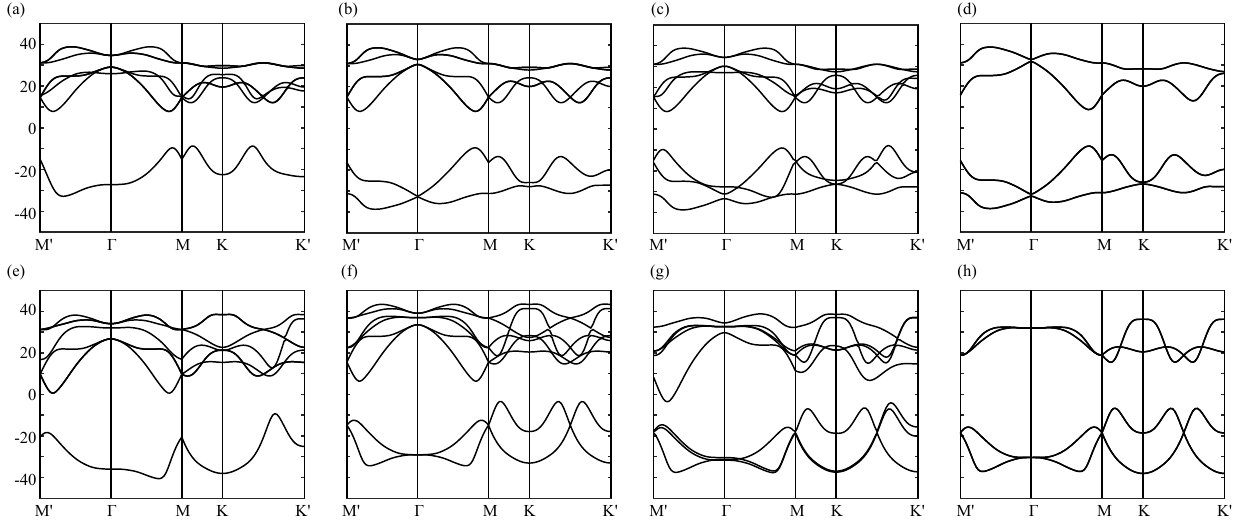}
    \caption{ The HF bands for ABABBC stacking order at integer fillings. From (a-d): $V_d=0\mrm{meV},\nu=-3,-2,-1,0$. From (e-h) $V_d=5\mrm{meV},,\nu=-3,-2,-1,0$. Here, we only show one type of groundstate, even for those fillings with different types of groundstates. The calculations are performed on $8\times8$ momentum lattice.}
    \label{fig:HFbands}
\end{figure*}

\begin{figure*}
    \centering
    \includegraphics[width=\linewidth]{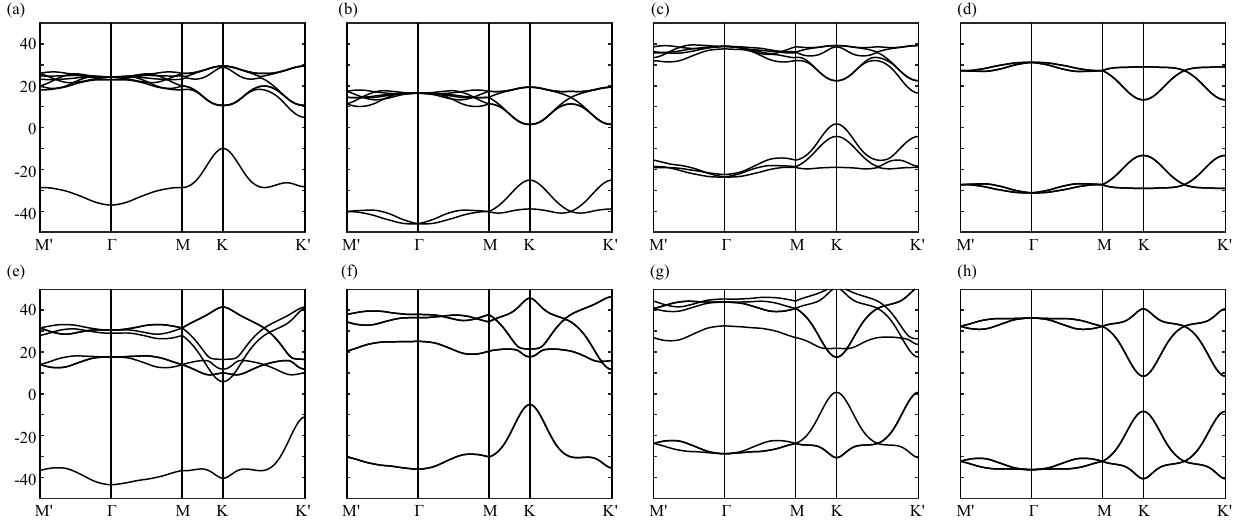}
    \caption{ The HF bands for ABABAB stacking order at integer fillings. From (a-d): $V_d=0\mrm{meV},\nu=-3,-2,-1,0$. From (e-h) $V_d=5\mrm{meV},,\nu=-3,-2,-1,0$. Here, we only show one type of groundstate, even for those fillings with different types of groundstates. The calculations are performed on $8\times8$ momentum lattice. }
    \label{fig:HFbands-ABABAB}
\end{figure*}

In this section we discuss the translational symmetric solutions at all integer fillings. As we will show in the next subsection, with the parameters we adopt, they are indeed the groundstates. With the aid of $\mathcal{P}_y$, which flips $\nu$ and $V_d$, we can only study $\nu\geq 0$ cases, and for ABABAB-stacking order, further with $C_{2x}$ symmetry which flips $V_d$, we can only study $\nu\geq 0,V_d\geq 0$ cases.

We first study ABABBC-stacking CTTBG in the absence of a displacement field. We find that in the absence of a vertical electric field, the groundstates at integer fillings are gapped and valley and/or spin-polarized states. Besides, among each valley the spin polarization at each $\kk$ points towards the same direction. We can classify the groundstates by their fillings of each valley and spin sector per Moir\'e unit cell, as shown in \cref{tab:GS}. We have applied SU(2)$_+\times$SU(2)$_-$ rotation to choose the spin polarization in $z$ direction. As an example, the degenerate groundstates at $\nu=0$ include
\begin{itemize}
    \item spin-polarized or spin-valley-locked states. These states break SU(2)$_+\times$SU(2)$_-$ symmetry but preserve a time reversal symmetry. As an example, one of these states has four spin-up electrons, and two lie in $\eta=+$ valley and two in $\eta=-$ valley. This state respects the spinless time reversal symmetry $T_0$. Other states are the SU(2)$_+\times$SU(2)$_-$ partners of this state and respect a certain combination of $T_0$ and SU(2)$_+\times$SU(2)$_-$ rotation.
    \item valley polarized state with four electrons polarized in one valley. This state preserves SU(2)$_+\times$SU(2)$_-$ but breaks TRS.
\end{itemize}
The two sets of states above are degenerate but are not related by symmetry. Instead, as detailed in \cref{app:degenerate}, some operations like acting $T_0$ to one spin sector of the order parameters can change one state to the other without changing HF energy functional. This degeneracy is the consequence of the HF approximation and should be lifted by fluctuations upon the HF saddle points. Previous studies \cite{zhang_correlated_2020,kwan_strong-coupling_2023} have also reported this degeneracy, and it is called "flavor permute symmetry" in Ref.\cite{kwan_strong-coupling_2023}.
For groundstates at $\nu=\pm 2$, we find that two electrons/holes tend to occupy different bands and maximize their total spin/valley polarization, which is similar to Hund's rule.

\begin{table}[h]
    \centering
    \begin{tabular}{c|c|c|c|c|c|c|c|c|c|c|c}\hline\hline
         \multirow{2}*{$\nu$} & \multicolumn{5}{c|}{FP (small $|V_d|$)} & \multicolumn{5}{c|}{LP (large $|V_d|$)} & $V_{d}^{(c)}$(meV)\\
         \cline{2-11}  & $\nu_{+\uparrow}$ & $\nu_{+\downarrow}$ & $\nu_{-\uparrow}$ & $\nu_{-\downarrow}$& $T$ & $\nu_{+\uparrow}$ & $\nu_{+\downarrow}$ & $\nu_{-\uparrow}$ & $\nu_{-\downarrow}$ & $T$\\\hline
         -3 & 1 & 0 & 0 & 0 & $\times$ & 1 & 0 & 0 & 0 &$\times$&\diagbox{}{}\\\hline
         \multirow{2}*{-2} & \multirow{2}*{2} & \multirow{2}*{0} &\multirow{2}*{0} &\multirow{2}*{0} &\multirow{2}*{$\times$} &  1 & 0 & 1 & 0 &\checkmark & -2, 2\\
         \cline{7-11}&&&&&&1 &1 &0&0 &$\times$&\\\hline
         \multirow{2}*{-1}&  2 &0&1 & 0 & $\times$&\multirow{2}*{1}& \multirow{2}*{1} & \multirow{2}*{1} &\multirow{2}*{0} & \multirow{2}*{$\times$}&-2, 2\\
         \cline{2-6}  &2& 1&0 & 0 &$\times$ &&&&&&\\\hline
         \multirow{2}*{0}&   2  & 0 &2 &0 &\checkmark&  \multirow{2}*{1} & \multirow{2}*{1} &\multirow{2}*{1} & \multirow{2}*{1} & \multirow{2}*{\checkmark}&-3, 3\\
         \cline{2-6}  &2 & 2 & 0 & 0 & $\times$ &&&&&&\\\hline\hline
    \end{tabular}
    \caption{The symmetry and occupation of groundstate at integer fillings, with or without electric field for ABABBC-stacking CTTBG. Notice that since they all preserve valley U(1), $T_0$ and $T_K$ make no difference so we just use $T$ to denote TRS. \label{tab:GS}}
\end{table}

\begin{table}[!ht]
    \centering
    \begin{tabular}{c|c|c|c|c}
    \hline\hline
        & $V_d$(meV) & 0 & 1 & $\geq$ 2 \\\hline 
         \multirow{3}*{-3} & $\mS_{V}$ & 1.99  & \multicolumn{1}{c}{} &\multirow{12}*{\phantom{1}\makecell[c]{Same as \\ LP in \\\cref{tab:GS}}}\phantom{1} \\ 
        \cline{2-3}~ & $\mS_{s+}$ & 0.62  \\ 
        \cline{2-3}~ & $\mS_{s-}$ & 0.63 \\
        \cline{1-4}\multirow{3}*{-2} & $\mS_{V}$ & 3.96 & 2.40 & ~ \\ 
        \cline{2-4}~ & $\mS_{s+}$ & 1.25 & 1.15 & ~ \\ 
        \cline{2-4} ~ & $\mS_{s-}$ & 1.25 & 1.15 & ~ \\ 
        \cline{1-4}\multirow{3}*{-1} & $\mS_{V}$ & 5.95 & 2.41 & ~ \\ 
        \cline{2-4}~ & $\mS_{s+}$ & 1.37 & 1.15 & ~ \\ 
        \cline{2-4}~ & $\mS_{s-}$ & 1.37 & 0.52 & ~ \\ 
        \cline{1-4}\multirow{3}*{0} & $\mS_{V}$ & 7.92 & 4.83 & ~ \\ 
        \cline{2-4}~ & $\mS_{s+}$ & 1.49 & 0.91 & ~ \\ 
        \cline{2-4}~ & $\mS_{s-}$ & \phantom{1}1.49\phantom{1} & \phantom{1}0.91\phantom{1} & ~ \\  \hline\hline
    \end{tabular}
    \caption{The symmetry-breaking strength of the groundstates at integer fillings for ABABAB-stacking CTTBG. $\mS_V,\mS_{s\pm}$ are the symmetry-breaking strength of valley U(1) and SU(2)$_\pm$ defined in \cref{sec:HF}.\label{tab:ABABAB-GS}}
\end{table}
The groundstates of ABABAB-stacking CTTBG at $V_d=0$ are completely different from that of ABABBC-stacking CTTBG, where the groundstates are IVC states at all integer fillings, which break U(1)$_V$ maximally as we find $\mS_{V}$ around 2$\nu$ at filling $\nu$. At fillings $\nu=0,\pm 2$ we find that the groundstates also preserve the Kramers TRS but do not preserve spinless TRS. In terms of the Chern basis introduced at the end of \cref{sec:singleparticle}, we find that IVC happens mainly in the bands with the same Chern number in ABABAB-stacking CTTBG, which is similar to IVC groundstates in TBG. In ABABBC-stacking CTTBG, all four sublattice-valley bands have different Chern numbers ($\pm2,\pm4$), thus, IVC order leads to vortices in Moir\'e BZ, which are disfavored \cite{bultinck_mechanism_2020,devakul_magic-angle_2023}.

We then move to the phase diagram at a finite displacement field. The groundstates of CTTBG for two stacking orders under the displacement field are also recorded in \cref{tab:GS,tab:ABABAB-GS}. We find that the displacement field will suppress the IVC order in ABABAB cases. Under a large enough displacement field, the groundstates of both stackings become the layer polarized states (LP). Applying a large enough vertical field separates the two active bands, where the upper/lower bands are mainly contributed by either top or bottom layers, depending on the direction of the electric field. Thus, one can obtain the groudstates with $V_d=5$meV in \cref{tab:GS,tab:ABABAB-GS} by assuming electrons are first layer-polarized and then minimized their interaction energy inside one layer. We do HF calculation at $V_d$ from $0$ to $9$ meV and from $-9$ to $9$ meV for each $1$ meV for ABABAB stacking order and ABABBC stacking order, respectively, and record the critical electric field $V_d^{(c)}$ where the groundstates switch between FP and IVC states and LP states in \cref{tab:GS,tab:ABABAB-GS}. We also plot the HF band at each integer filling in \cref{fig:HFbands,fig:HFbands-ABABAB}.

\subsection{Charge density wave states }

\begin{table}[htbp]
    \centering
    \begin{tabular}{c|c|c|c|c|c|c|c|c|c|c|c}\hline\hline
         \multirow{2}*{$\nu$} & \multicolumn{5}{c|}{FP(small $|V_d|$)} & \multicolumn{5}{c|}{LP(large $|V_d|$)} & \multirow{2}*{\makecell{$V_d^{(c)}$\\(meV)}} \\
         \cline{2-11}  & $\nu_{+\uparrow}$ & $\nu_{+\downarrow}$ & $\nu_{-\uparrow}$ & $\nu_{-\downarrow}$& $T$ & $\nu_{+\uparrow}$ & $\nu_{+\downarrow}$ & $\nu_{-\uparrow}$ & $\nu_{-\downarrow}$ & T\\\hline
         -3.5 & 0.5 & 0 & 0 & 0 & $\times$ & 0.5 & 0 & 0 & 0 &$\times$ &\diagbox{}{}\\\hline
         \multirow{2}*{-2.5} & \multirow{2}*{1.5} & \multirow{2}*{0} &\multirow{2}*{0} &\multirow{2}*{0} &\multirow{2}*{$\times$} &  1 & 0 & 0.5 & 0 &$\times$&-2, 2\\
         \cline{7-11}&&&&&&1 &0.5 &0&0 &$\times$&\\\hline
         \multirow{2}*{-1.5}&  2 &0&0.5 & 0 & $\times$&1& 0.5 & 1 &0 & $\times$&-2, 2\\
         \cline{2-11}  &2& 0.5&0 & 0 &$\times$ &1&1&0.5&0&$\times$&\\\hline
         \multirow{2}*{-0.5}&   2  & 0 &1.5 &0 &\checkmark&  \multirow{2}*{1} & \multirow{2}*{1} &\multirow{2}*{1} & \multirow{2}*{0.5} & \multirow{2}*{$\times$}&-2, 2\\
         \cline{2-6}  &2 & 1.5 & 0 & 0 & $\times$ &&&&&&\\\hline\hline
    \end{tabular}
    \caption{The symmetry and occupation of groundstate at half-integer fillings, with or without electric field for ABABBC-stacking CTTBG.  \label{tab:ABABBC-GS-half}}
\end{table}

\begin{table}[!ht]
    \centering
    \begin{tabular}{c|c|c|c|c}
    \hline\hline
        & $V_d$(meV) & 0 & 1 & $\geq$ 2 \\\hline 
         \multirow{3}*{-3.5} & $\mS_{V}$ & 1.0  & \multicolumn{1}{c}{} &\multirow{12}*{\phantom{1}\makecell[c]{Same as \\ LP in \\\cref{tab:ABABBC-GS-half}}}\phantom{1} \\ 
        \cline{2-3}~ & $\mS_{s+}$ & 0.31  \\ 
        \cline{2-3}~ & $\mS_{s-}$ & 0.31 \\
        \cline{1-4}\multirow{3}*{-2.5} & $\mS_{V}$ & 2.98 & 0.82 & ~ \\ 
        \cline{2-4}~ & $\mS_{s+}$ & 0.94 & 1.09 & ~ \\ 
        \cline{2-4} ~ & $\mS_{s-}$ & 0.94 & 0.52 & ~ \\ 
        \cline{1-4}\multirow{3}*{-1.5} & $\mS_{V}$ & 4.95 & 2.39 & ~ \\ 
        \cline{2-4}~ & $\mS_{s+}$ & 1.32 & 1.15 & ~ \\ 
        \cline{2-4}~ & $\mS_{s-}$ & 1.29 & 0.83 & ~ \\ 
        \cline{1-4}\multirow{3}*{-0.5} & $\mS_{V}$ & 6.94 & 3.24 & ~ \\ 
        \cline{2-4}~ & $\mS_{s+}$ & 1.43 & 0.96 & ~ \\ 
        \cline{2-4}~ & $\mS_{s-}$ & \phantom{1}1.43\phantom{1} & \phantom{1}0.65\phantom{1} & ~ \\  \hline\hline
    \end{tabular}
    \caption{The symmetry-breaking strength of the groundstates at half-integer fillings for ABABAB-stacking CTTBG. $\mS_V,\mS_{s\pm}$ are the symmetry-breaking strength of valley U(1) and SU(2)$_\pm$ defined in \cref{sec:HF}.\label{tab:ABABAB-GS-half}}
\end{table}

\begin{figure}
    \centering
    \includegraphics[width=\linewidth]{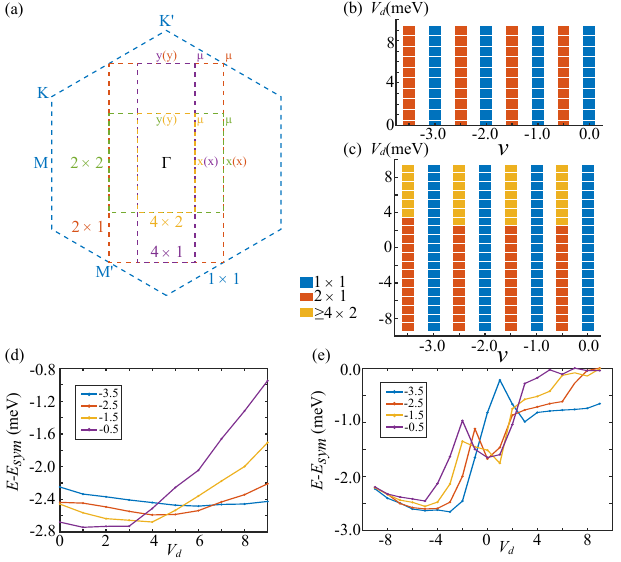}
    \caption{(a) The folded Brillouin zone considered in this work. (b)(c) The CDW grounstates at all integer and non-integer fillings for ABABAB and ABABBC stacking orders, respectively. (d)(e) Energy difference of $2\times1$ charge density wave states and translational symmetric states at half-integer fillings for ABABAB and ABABBC stacking orders, respectively. Notice that we only plot $V_d>0$ cases for ABABAB stacking order due to $C_{2x}$ symmetry.}
    \label{fig:CDW}
\end{figure}

\begin{figure}
    \centering
    \includegraphics[width=\linewidth]{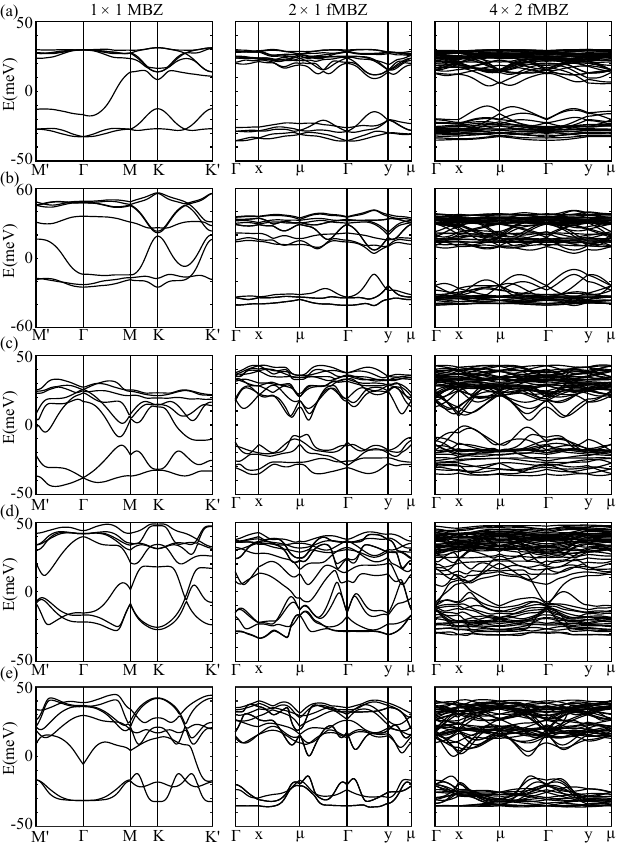}
    \caption{(a)-(e) The HF bands at $\nu=-1.5$ for ABABAB stacking CTTBG under $V_d=0,5$meV amd ABABBC stacking CTTBG under $V_d=0,5,-5$meV, respectively. The bands are calculated in $1\times1$,$2\times1$ and $4\times2$ fMBZ from left to right in each row. The momentum lattice used is $8\times8$.}
    \label{fig:CDWband}
\end{figure}

In this section, we consider possible translation-symmetry-breaking solutions with fMBZ as shown in \cref{fig:CDW}(a) and compare their energies to determine the groundstates. Importantly, only after checking that further breaking translation symmetry upon a state will yield the same solution can we verify this state as a groundstate. For example, solutions in $2\times2$ fMBZ are groundstates when the ones in $4\times2$ fMBZ are the same as them, but since we calculate $4\times2$ fMBZ at most if the solutions in $4\times2$ fMBZ have the lowest energy, the true groundstates are not yet determined since other translation-symmetry-breaking states, \textit{e.g.}, $4\times4$ density wave states may be the groundstates. 

Using the criterion above, we finally label the fMBZ that groundstate corresponds to with different colors under different displacement fields in \cref{fig:CDW}(b)(c). We find that the groundstates are translational symmetric states at integer fillings and density wave states at half-integer fillings. For ABABAB stacking CTTBG, the density wave states are all $2\times 1$ stripe states while for ABABBC stacking CCTBG, a large enough positive $V_d$ will prefer further translation-symmetry-breaking than $2\times 1$ stripe states. 
We plot the band structures for $\nu=-1.5$ as an example in \cref{fig:CDWband}. 
The translational symmetric solutions at half-integer fillings are always gapless as they have partially filled HF bands, as shown in the left panel of \cref{fig:CDWband}(a-e), while breaking translation symmetry tends to open a gap which lowers the total energy. 
It turns out that if a HF solution is already gapped in $2\times1$ fMBZ, it will be the groundstate, \textit{e.g.}, the $\nu=-1.5$ states for ABABAB stacking at $V_d=0,5$ meV and ABABBC stacking at $V_d=0,-5$meV as shown in the middle panel of \cref{fig:CDWband}(a)(b)(c)(e). Otherwise, further translation-symmetry-breakings are preferred. 
Especially, here we find that all these states remain gapless even at $4\times2$ fMBZ, \textit{e.g.}, the $\nu=-1.5$ states for ABABBC stacking at $V_d=5$meV as shown in \cref{fig:CDWband}(d), which suggests that further translation-symmetry-breaking may happen. 
We tabulate the groundstates at half-integer fillings for ABABBC and ABABAB stacking in \cref{tab:ABABBC-GS-half,tab:ABABAB-GS-half}. We emphasize that \cref{tab:ABABBC-GS-half} is valid only when the groundstates are the solution in $1\times1$ and $2\times1$ fMBZ, since for those cases which have lowest energy in $4\times2$ fMBZ calculation, we do not know whether we have found the groundstate and plot the energy difference of $2\times1$ CDW states and translational symmetric states for two stacking order as a function of electric field in \cref{fig:CDW}(d)(e).  The CDW orders are tuned differently under the electric field for the two stacking orders. 

Compared to TBG, where the CDW states are rarely reported except for $\nu=3$ \cite{xie_phase_2023}, CDW states are common as the groundstate in CTTBG, regardless of the stacking order. We attribute this to the distinct charge density profile of the flat band of CTTBG and TBG. As shown in \cref{fig:density}, the distribution of flat band electrons in CTTBG is more spread over the Moir\'e unit cell, while in TBG, flat band electrons are highly concentrated at the AA-stacking region. The highly localized orbitals in TBG  allow one to construct an effective model when nonlocal interactions are negligible, and the correlation effect mostly originates from the on-site Hubbard interaction \cite{song_magic-angle_2022,zhou_kondo_2024,wang_molecular_2024}. In contrast, the nonlocal Coulomb interaction plays a much more crucial role in CTTBG, which avoids electrons occupying neighboring sites and explains the existence of CDW states in CTTBG.

\section{Summary and Discussion}
In this work, we study the single-particle physics and mean field phase diagram of CTTBG of two stacking orders, namely ABABAB and ABABBC. We find that there are a pair of flat bands with nontrivial topology at $\theta=1.70^\circ$ in these two cases. For ABABBC, we predict the valley Hall effect at fillings $\pm 4$. We further perform HF calculation to determine the phase diagram of CTTBG and find that the groundstates are flavor polarized states for ABABBC stacking order and IVC states for ABABAB stacking order at a small displacement field. Both become layer-polarized states at a large enough displacement field. We find that the groundstates at half-integer fillings are charge density wave states. For ABABAB stacking among a range of displacement fields, the groundstates are always $2\times1$ stripe states. For ABABBC stacking the groundstates are also $2\times1$ stripe states at a small displacement field, and a larger displacement will possibly favor further translation-symmetry-breakings, depending on the direction of the displacement field and filling. We demonstrate that the observed CDW states can originate from the strong Coulomb interaction compared to the small kinetic energy and the real space distribution of the flat band electrons.

We focus on the CDW caused by electron-electron interaction in this work. However, we should point out that this is only a candidate for the mechanism of the CDW states observed in CTTBG \cite{wang_correlated_2024}, and there are other possibilities like electron-phonon coupling. Further experimental and theoretical research is needed to determine the origin of CDW states in CTTBG. 

\begin{acknowledgements}
Z.-D. S., Y.-J. W. and G.-D. Z. were supported by National Natural Science Foundation of China (General Program No. 12274005), National Key Research and Development Program of China (No. 2021YFA1401900), and Innovation Program for Quantum Science and Technology (No. 2021ZD0302403). X. L. acknowledges support from the National Key R\&D Program (Grant nos. 2022YFA1403500/02) and the National Natural Science Foundation of China (Grant Nos. 12274006 and 12141401).
\end{acknowledgements}
\bibliography{refs.bib}
\appendix

\section{More details about Model and convention\label{app:model}}

In this section, we derive the continuum model \cite{bistritzer_moire_2011,khalaf_magic_2019} for the chirally twisted triple bilayer graphene (CTTBG) detailedly. As in the main text, the rotation angle of the three bilayers $i=1,2,3$ are $(\theta,0,-\theta)$ from top to bottom. Like the case in twist bilayer graphene (TBG) \cite{bistritzer_moire_2011}, the low energy states in CTTBG are also contributed by the original low energy states near $\mathbf{K},\mathbf{K}'$ point of the untwisted unit, which is the single-layer graphene in TBG and Bernal-stacking bilayer graphene here. We choose the Bloch basis 
\begin{equation}
    c_{i,\pp,a}=\frac{1}{\sqrt{N}}\sum_{\RR_i}e^{-i\pp\cdot R_{\theta_i}(\RR_i+\ttau_a+\dd_i)}c_{i,\RR_i,a}
\end{equation}
where $\RR_i$ are lattice vectors of the $i$-th bilayers, $a=l\alpha$ in order (1A,1B,2A,2B) labels the layer index $l$ in each bilayer graphene and sublattice $\alpha$ (1 is the top layer and 2 is the bottom layer), $\ttau_a$ is the atom position inside the unit cell while $\ttau_a=\ttau_\alpha+\ttau_l$ where $\ttau_\alpha,\ttau_l$ define the position of sublattice and monolayer graphene inside bilayer. Besides, we define $\dd_i$ to be the displacement for the $i$-th bilayer, and if $\dd_i=0$, the AB stacking region of the bilayer graphene is located in the origin. 
$\RR_i,\ttau_a,\dd_i$ are all measured in the reference frame of each graphene bilayer, which require rotations to be their actual value in the laboratory frame. $\pp$ is defined in the laboratory frame.
We choose $\ttau_{1A}-\ttau_{1B}=\ttau_{2A}-\ttau_{2B}=-\frac{a_0}{\sqrt{3}}(0,1)$ in each layer and for Bernal stacking we have $\ttau_{2B}=\ttau_{1A}$ where $a_0=2.46$ \AA. 

\begin{figure*}
    \centering
    \includegraphics[width=\linewidth]{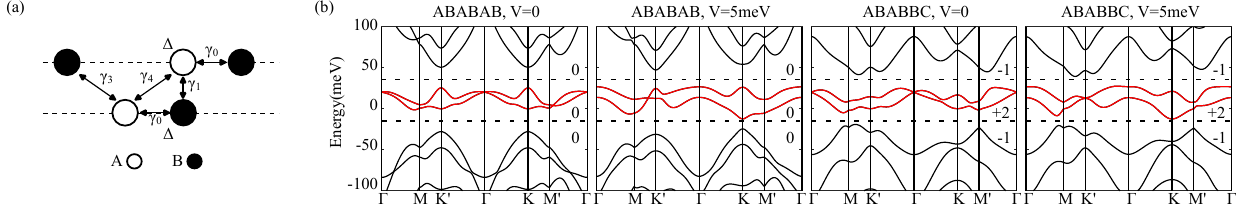}
    \caption{(a) The definition of tight binding parameters in Bernal bilayer graphene. (b) The single particle band when $\gamma_3,\gamma_4,\Delta$ are taken into consideration for ABABAB stacking CTTBG under $V_d=0,5$meV and ABABBC stacking CTTBG under $V_d=0,5$meV from left to right. The numbers mark the total Chern number of the remote conductance bands, the flat bands and the remote valence bands from top to bottom, respectively.}
    \label{fig:singleparticle-nnn}
\end{figure*}

We adopt the model for each Bernal-stacking bilayer graphene from \cite{jung_accurate_2014,lee_theory_2019} where the parameters $\gamma_{0,1,2,3,4}$ and $\Delta$ are defined in \cref{fig:singleparticle-nnn}. In the main text we keep only $\gamma_{0,1}$ and drop the other terms which break the particle hole symmetry $\mathcal{P}_y$.The Hamiltonian for the middle layer is
\begin{equation}
    H^{(22)}(\kk)=\left(\begin{array}{cccc}
         \Delta & -\gamma_0 f(\kk) &\gamma_4 f^*(\kk) & \gamma_1  \\
         -\gamma_0 f^*(\kk) & 0 &\gamma_3 f(\kk) & \gamma_4 f^*(\kk)  \\
         \gamma_4 f(\kk) & \gamma_3 f^*(\kk) &0 & -\gamma_0 f(\kk)  \\
         \gamma_1 & \gamma_4 f(\kk) &-\gamma_0 f^*(\kk) & \Delta  
    \end{array}
    \right)
\end{equation}
where $f(\kk)=\sum_{j=1}^3 e^{-i\kk\bm{\delta}_j}$ and $\bm{\delta}_j=C_{3z}^{j-1}(\ttau_{1A}-\ttau_{1B}),j=1,2,3$ are the vectors from B site to nearest neighbor A sites.
The Hamiltonian for other bilayers could be obtained by 
\begin{equation} 
    H^{(11)}(\kk)=H^{(22)}(R_\theta^{-1}\kk),H^{(33)}(\kk)=H^{(22)}(R_\theta\kk)\, .
\end{equation}
We have $f(\eta\KK+\kk)\approx \frac{\sqrt{3}}{2}a_0(\eta \kk_x+i\kk_y)$ for small $|\kk|$. We denote $h^{(\eta)}(\kk)$ as the low energy Hamitonian near  $\eta \KK$ such that $h^{(\eta)}(\kk)=H^{(22)}(\eta\KK+\kk)$.
We then consider hopping between different bilayer graphenes. The hopping term in the two-center, tight-binding approximation is
\begin{align}
    \hat{T}^{ij} &=\sum_{\RR,\RR',\langle aa'\rangle} t\big[R_{\theta_i}(\RR+\ttau_a+\dd_i)\nonumber \\
    &-R_{\theta_j}(\RR'+\ttau_{a'}+\dd_j)\big]c^\dagger_{i\RR a}c_{j\RR' a'}
\end{align}
where here $\langle aa'\rangle$ means $a$ in the $i$-th bilayer and $a'$ in the $j$-th bilayer are in adjacent layers, which should not be confused with nearest neighbor sites, and $i,j$ are also restricted to adjacent layers.
We expand $t(\rr)=\frac{1}{N\Omega}\sum_{\qq}e^{i\qq\cdot\rr}t_{\qq}$ where $\qq$ goes over all momentum space, yielding
\begin{align}
    \hat{T}^{ij}=\sum_{\substack{\pp\pp'\GG\GG'}}\sum_{\langle aa'\rangle} \frac{t_{\pp+R_{\theta_i}\GG}}{\Omega}\delta_{\pp+R_{\theta_i}\GG,\pp'+R_{\theta_j}\GG'} \nonumber \\
    e^{i\GG\cdot(\ttau_a+\dd_i)-i\GG'\cdot(\ttau_{a'}+\dd_j)} c^\dagger_{i\pp a}c_{j\pp' a'} \label{eq:T-gbasis}
\end{align}
where $\GG,\GG'$ sum over unrotated reciprocal lattice, and $\pp,\pp'$ sum over the BZ for the $i,j$-th bilayers, denoted as BZ$_i$ with reciprocal lattice $R_{\theta_{i,j}}\GG$, respectively.

We first consider only the low energy theory in $\eta=+$ valley  where $\pp,\pp'$ are close to $\KK=\frac{4\pi}{3a_0}(1,0)$ and expand $\pp=R_{\theta_i}\KK+\delta \pp$ and relabel $\delta \pp$ as $\pp$. 
Then the Hamiltonian of the intralayer part is just $H^{ii}(R_{\theta_i}\KK+ \pp) = h^{+}(R_{\theta_i}^{-1} \pp)\approx h^{+}(\pp)$ when $\theta$ is small\cite{song_all_2019}. 
For the interlayer part, since $t_{\qq}$ decays exponentially as $|\qq|$ increases, we keep only $\GG=\GG_{1,2,3}$ where $\GG_1=0,\GG_2=C_{3z}\KK-\KK,\GG_3=C_{3z}^2\KK-\KK$, which corresponds to the three largest hopping $\approx t_{\KK},t_{C_{3z}\KK},t_{C_{3z}^2\KK}$. By $C_{3z}$ they are the same and we denote them as $w\Omega$. Taking into the lattice relaxation effect, the hopping is not the same for AA and AB hopping so we use $w_{aa'}$ instead of a single $w$.   Furthermore, As $\pp,\pp'$ are both close to  $\KK^+$ the $\delta$ function in \cref{eq:T-gbasis} also gives $\GG=\GG'$. So the Moir\'e hopping term in $\eta=+$ valley is
\begin{align}
    & \hat{T}^{ij}_{+} = \sum_{\pp\pp'n}\sum_{\langle aa'\rangle}w_{aa'}e^{i\GG_n(\ttau_a+\dd_i-\ttau_{a'}-\dd_j)}c^\dagger_{i\pp a}c_{j\pp-\qq^n_{ij}a'} \label{eq:MoireHopping} 
\end{align}
which can also be explicitly written down as
\begin{align}
    & \hat{T}^{i,i+1}_{+} =  \sum_{\pp\pp'n}\sum_{aa'}\left[\left(\begin{array}{cc}
        0 & 0 \\
        1 & 0 
    \end{array}\right)\otimes T_n\right]_{aa'}\nonumber \\
    & \qquad\qquad e^{i\GG_n(\bm{\delta}_1+\dd_i-\dd_{i+1})}c^\dagger_{i,\pp, a}c_{i+1,\pp-\qq^n_{i,i+1},a'} \label{eq:MoireHopping1} \\
    &T_n = w_0\sigma_0 + w_1\sigma_x\cos \frac{2\pi(n-1)}{3}  + w_1\sigma_y \sin\frac{2\pi(n-1)}{3} 
\end{align}
where $\qq^n_{ij}=R_{\theta_j}\GG_n-R_{\theta_i}\GG_n$, satisfying $\qq^n_{ij}=C^{(n-1)}_{3z}\qq^1_{ij}$ and $w_0,w_1$ are interlayer AA and AB couplings. In principle, $\qq^n_{12}=R_{\theta}\qq^n_{23}$ instead of $\qq^n_{23}$, which leads to the moir\'e of moir\'e structure. Here, we neglect this effect and let $\qq^n_{12}=\qq^n_{23}\equiv \qq_n$, with $\qq_1=k_\theta(0,-1)$, where $k_\theta=2|\KK|\sin\frac{\theta}{2}$. 

\cref{eq:MoireHopping} indicates that an electron with momentum $\pp$ in layer $i$ can couple to electron in the same bilayer with momentum $\pp+\bb_1,\pp+\bb_2$ via Moir\'e coupling with other bilayers, where $\bb_1=\qq_2-\qq_1=k_\theta(\frac{\sqrt{3}}{2},\frac{3}{2}),\bb_2=\qq_3-\qq_1=k_\theta(-\frac{\sqrt{3}}{2},\frac{3}{2})$, which form a basis of Moir\'e reciprocal lattice.
We then rewrite $\pp$ in the $i$-th layer as $\kk-\QQ_i$, where $\kk$ lies in the first Moir\'e Brillouin zone (MBZ) and $\QQ_i$ run over the set $\mathcal{Q}_i$, which is formed by $-\qq_{1},\mathbf{0},\qq_{1}$ plus Moir\'e reciprocal lattice for $i=1,2,3$ respectively. Then the Moir\'e hopping term becomes
\begin{align}
    &\hat{T}^{i,i+1}_+ =  \sum_{\substack{\kk\in\mathrm{MBZ}\\\QQ\in\mathcal{Q}_i,\QQ'\in\mathcal{Q}_{i+1}}}\sum_{aa'} [T^{i,i+1}_{+}(\kk)]_{\QQ,a;\QQ',a'} c^\dagger_{\kk\QQ a}c_{\kk\QQ'a'} \label{eq:MoireHopping-MBZ} \\
    &[T^{i,i+1}_{+}(\kk)]_{\QQ,a;\QQ',a'}  = \sum_n \left[\left(\begin{array}{cc}
        0 & 0 \\
        1 & 0 
    \end{array}\right)\otimes T_n\right]_{aa'}  \nonumber \\
    & \qquad\qquad\qquad\qquad\qquad e^{i\GG_n(\bm{\delta}_1+\dd_i-\dd_{i+1})}\delta_{\QQ+\qq_n,\QQ'}
\end{align}
where we have neglected the $\theta$ dependence in $H^{ii}$.

We could see the explicit dependence of the Moire potential over $\dd$ by performing a gauge transformation $c_{j}\to c_{j}e^{-i\KK\cdot (\dd_j-j\bm{\delta}_1)}$ and writing \cref{eq:MoireHopping} in real space by $c_{i\pp a}=\int {\rm d}\rr^2 e^{-i\pp\cdot\rr}c_{i\rr a}$ 
\begin{align}
    \hat{T}^{i,i+1}_+ &=  \sum_{n}\sum_{aa'}\int{\rm d}^2\rr\left[\left(\begin{array}{cc}
    0 & 0 \\
    1 & 0 
    \end{array}\right)\otimes T_n\right]_{aa'}\nonumber \\
    & \qquad e^{iC_{3z}^{n-1}\KK\cdot(\bm{\delta}_1+\dd_i-\dd_{i+1})+iq_n\cdot\rr}c^\dagger_{i,\rr, a}c_{i+1,\rr,a'}  \\
    & = \sum_{n}\sum_{aa'}\int{\rm d}^2\rr\left[\left(\begin{array}{cc}
    0 & 0 \\
    1 & 0 
    \end{array}\right)\otimes T_n\right]_{aa'}\nonumber \\
    & \qquad e^{i\qq_n\cdot[\rr-\hat{z}\times(\bm{\delta}_1+\dd_i-\dd_{i+1})/(2\sin\frac{\theta}{2})]}c^\dagger_{i,\rr, a}c_{i+1,\rr,a'}  \label{eq:Moire-realspace}
\end{align}
where we have made use of $C^{n-1}_{3z}\KK\cdot \dd = \frac{1}{2\sin\frac{\theta}{2}}(\hat{z}\times \qq_n)\cdot \dd =  \frac{1}{2\sin\frac{\theta}{2}}(\dd\times\hat{z})\cdot \qq_n  $. In \cref{eq:Moire-realspace} one can see that the Moir\'e pattern is shifted by $\hat{z}\times(\dd_i-\dd_{i+1})/(2\sin\frac{\theta}{2})$ due to the displacement.

We then add the Hamiltonian of valley $\eta=-$. By acting TRS on Hamiltonian in $\eta=+$ valley, we have $h^{-}_{aa'}(\kk-\QQ) = h^{+,*}_{aa'}(-\kk+\QQ)$ and $[T^{i,i+1}_{-}(\kk)]_{\QQ, a;\QQ' a'} = [T^{i,i+1}_{+}(-\kk)]^*_{-\QQ,a;-\QQ',a'} $. The BM Hamiltonian is then
{\small
\begin{align}
    \hat{H}_{BM}=\sum_{\kk\in {\rm MBZ}} \sum_{\QQ aa'} h^{\eta}_{aa'}(\kk-\QQ) c^\dagger_{\kk\QQ a\eta}c_{\kk\QQ a'\eta}+ \sum_{\eta}\sum_{i} \hat{T}^{i,i+1}_{\eta}
\end{align}
}
Finally, we add the vertical electric field and obtain the total non-interacting Hamiltonian $\hat{H}_0 = \hat{H}_{BM} + \hat{H}_{V}$, where
\begin{align}
    \hat{H}_V = V_d\sum_{\kk\in {\rm MBZ}}\sum_{\QQ aa'}\left(l_{\QQ a}-\frac{7}{2}\right)c^\dagger_{\kk\QQ a\eta}c_{\kk\QQ a\eta'}
\end{align}
For $c^\dagger_{\kk\QQ a\eta},\QQ\in Q_i$ which comes from $a$ sublattice and the $i$-th bilayer we label the monolayer graphene it belongs to as $l_{\QQ a}$.

We end this section with a discussion on the influence of the hopping terms $\gamma_{3,4}$ and $\Delta$. We plot the single particle bands in \cref{fig:singleparticle-nnn}(b). Compared to \cref{fig:singleparticle}(c) where the two flat bands are separated by the displacement, here there is not a gap between them, so the Chern number of each band is not well defined. However, we find that the total Chern number of the flat bands and the remote conduction/valence bands are separately the same as those without $\gamma_{3,4},\Delta$.

\section{Projected model and Hatree-Fock calculation \label{app:proj-HF}}
\subsection{Gauge fixing}
For later convenience, we fix the gauge choice of flat band operators. We expand the electron operators in the eigenstate of $\hH_0$ into the original basis as $c^\dagger_{\kk,m,\eta,s}=\sum_{\QQ,a} u^{(\eta)}_{\QQ a,m}(\kk)c^\dagger_{\kk,\QQ,a,\eta,s}$, where $m=\pm1$ labels the flat bands and $|m|>1$ are remote valence/conduction band. We choose the periodical gauge $c^\dagger_{\kk+\GG,m,\eta,s}=c^\dagger_{\kk,m,\eta,s}$ so $u^{(\eta)}_{\QQ a,m}(\kk+\GG)=u^{(\eta)}_{\QQ-\GG a,m}(\kk)$, and the electrons in two valley are related by $c^\dagger_{\kk,m,\eta,s} = T_0 c^\dagger_{-\kk,m,-\eta,s} T_0^{-1}$ so $u^{(\eta)}_{\QQ a,m}(\kk)=u^{(-\eta)*}_{-\QQ a,m}(-\kk)$.
\subsection{Projected model}
The interaction Hamiltonian we consider are
\begin{align}
    \hH_I&=\frac{1}{2\Omega_{tot}}\sum_{\qq\in \MBZ}\sum_{\GG}V(\qq+\GG)\delta\rho_{\qq+\GG}\delta\rho_{-\qq-\GG},\label{eq:Hint}
\end{align}
where $V(\qq+\GG)$ is the Fourier component of the electron-electron interaction, which we take to be the double-gate-screened Coulomb interaction $V(\qq)=\frac{e^2}{2\epsilon|\qq|}\tanh{\frac{\xi |\qq|}{2}}$, and
{
\begin{align}
    \delta\rho_{\qq+\GG}&=\sum_{\eta\alpha s}\sum_{\kk\in\mrm{MBZ}}\sum_{\QQ\in\mathcal{Q}} \nonumber \\
    &\left(c^\dagger_{\kk+\qq,\QQ-\GG,a,\eta,s}c_{\kk,\QQ,a,\eta,s} - \frac{1}{2}\delta_{\qq,\oo}\delta_{\GG,\oo}\right) \label{eq:rho}
\end{align}
}
is the Fourier component of the density operator relative to CNP.

Hereafter we write the total model $\hH_0+\hH_I$ on the eigenstate basis of $\hH_0$

\begin{align}
    \hH_0&=\sum_{\kk\in \MBZ}\sum_{\eta,m,s}\epsilon^{(\eta)}_{m}(\kk)c^\dagger_{\kk m\eta s}c_{\kk m\eta s}\label{eq:H0-band-basis}
\end{align}
and $\hH_I$ keeps the form in \cref{eq:Hint} with density operator becoming
{\small 
\begin{align}
    \delta\rho_{\qq+\GG}=\sum_{\kk,\eta,s}\sum_{mn}M^{(\eta)}_{mn}(\kk;\qq+\GG)\left( c^\dagger_{\kk+\qq m\eta s}c_{\kk n\eta s}-\frac{1}{2}\delta_{\qq,0}\delta_{mn}\right)\label{eq:density}
\end{align}
}
\begin{align}
    M^{(\eta)}_{mn}(\kk;\qq+\GG)=\sum_{\QQ a}u^{(\eta)*}_{\QQ-\GG a,m}(\kk+\qq) u^{(\eta)}_{\QQ a,n}(\kk) \label{eq:M}
\end{align}

Substitute \cref{eq:density} into \cref{eq:Hint} we have
\begin{widetext}

{\small
\begin{equation}
    \hH_I=\frac{1}{2\Omega_{tot}}\sum_{\kk\kk'\qq\in \MBZ}\sum_{ss',\eta\eta'}\sum_{mm'nn'}U^{(\eta\eta')}_{mn,m'n'}(\qq;\kk,\kk')\left( c^\dagger_{\kk+\qq m\eta s}c_{\kk n\eta s}-\frac{1}{2}\delta_{\qq,0}\delta_{mn}\right)\left( c^\dagger_{\kk'-\qq m'\eta' s'}c_{\kk' n'\eta' s'}-\frac{1}{2}\delta_{\qq,0}\delta_{m'n'}\right) \label{eq:Hint-band-basis}
\end{equation}
}
\begin{align}
    U^{(\eta\eta')}_{mn,m'n'}(\qq;\kk,\kk')&=\sum_{\GG}V(\qq+\GG)M^{(\eta)}_{mn}(\kk,\qq+\GG)M^{(\eta')}_{m'n'}(\kk',-\qq-\GG) \label{eq:U}
\end{align}
\end{widetext}

To write down the effect model on the flat bands, we only keep $m,n$ for the flat bands in \cref{eq:H0-band-basis,eq:Hint-band-basis}. One should notice that this is not directly projecting \cref{eq:Hint-band-basis} on the flat band, where HF potential from the remote bands should also be included. In fact, the actual interaction term we consider includes not only  \cref{eq:Hint} but also a subtraction term in order to avoid double counting of the interactions included in DFT bands and HF calculation, as done in previous works in Moir\'e materials. Readers can refer to the appendices of Ref.\cite{parker_field-tuned_2021,kwan_moire_2023} which discuss this in detail. In short, the subtraction lets the HF contribution of $\hH_I$ to the energy bands vanish at a certain reference density. One can check that our procedure here is just the infinite temperature subtraction scheme in Ref.\cite{parker_field-tuned_2021}, where the Hatree-Fock decomposition vanishes when the active bands are all half-filled, the remote valence bands are fully filled and remote conductance bands are empty.

We also summarized the symmetry, hermicity, and periodicity of $M$ and $U$ here, which can be used to check the correctness of numerical results. 
\begin{itemize}
    \item By time reversal symmetry we have $M^{(-\eta)}_{mn}(\kk;\qq+\GG)=M^{(\eta)*}_{mn}(-\kk;-\qq-\GG)$. 
    \item With the periodicity in reciprocal space we have $M^{(\eta)}_{mn}(\kk+\GG';\qq+\GG)=\sum_{\QQ a}u^{(\eta)*}_{\QQ-\GG-\GG' a,m}(\kk+\qq) u^{(\eta)}_{\QQ-\GG' a,n}(\kk)=M^{(\eta)}_{mn}(\kk;\qq+\GG)$. $U^{(\eta\eta')}_{mn,m'n'}(\qq;\kk,\kk')$ is periodical in reciprocal space for both $\qq,\kk,\kk'$.
    \item By hermicity we have $M^{(\eta)*}_{mn}(\kk;\qq+\GG)=M^{(\eta)}_{nm}(\kk+\qq+\GG;-\qq-\GG) = M^{(\eta)}_{nm}(\kk+\qq;-\qq-\GG)$,$U^{(\eta\eta')*}_{mn,men's}(\qq;\kk,\kk')=U^{(\eta'\eta)}_{n'm',nm}(\qq;\kk'-\qq,\kk+\qq)$
    \item Since $V(\qq)=V(-\qq)$, we have $U^{(\eta\eta')}_{mn,m'n'}(\qq;\kk,\kk')=U^{(\eta'\eta)}_{m'n',mn}(-\qq;\kk',\kk)$ as we have summed over infinite $\GG$.
\end{itemize}  

Notice that some properties above depend on summing over infinite $\QQ$ in \cref{eq:M} and infinite $\GG$ in \cref{eq:U}, so in the numerical calculation where only finite $\QQ,\GG$ are used they are not exactly satisfied but the deviation will decay as the cutoff increases.

\subsection{Hatree-Fock method}
We perform the Hatree-Fock calculation in momentum space on $\hH_0+\hH_I$. Spontaneous breaking of the translation symmetry is allowed, and the order parameters are $O_{b m\eta s,b' n\eta' s'}(\kk)=\langle c^\dagger_{\kk+\QQ_b,m,\eta,s}c_{\kk+\QQ_{b'}, n,\eta,s'}\rangle-\frac{1}{2}\delta_{bb'}\delta_{mn}\delta_{ss'}\delta_{\eta\eta'}$.
The mean field Hamiltonian depends only on $O_{b m\eta s,b' n\eta' s'}(\kk)$:
\begin{widetext}
\begin{align}
    \hH_{I}^{MF}=\frac{1}{\Omega_{tot}}\sum_{\kk\in \fMBZ}\sum_{bb'}\sum_{m\eta s,n\eta's'} (H^{MF}_{I})_{b m\eta s,b' n\eta' s'}(\kk) c^\dagger_{\kk+\QQ_b m\eta s}c_{\kk+\QQ_{b'} n\eta' s'} \label{eq:HMF-fold}
\end{align}
where 
\begin{align}
    (H^{MF}_{I})_{b m\eta s,b' n\eta' s'}(\kk)&=\sum_{b''b'''}\sum_{m'n'}\sum_{\kk'}\Big[\delta_{ss'}\delta_{\eta\eta'}\sum_{\eta''s''}  U^{(H,\eta\eta'')}_{mnbb';m'n'b''b'''}(\kk,\kk') O_{b'' m'\eta'' s'',b''' n'\eta'' s''}(\kk') \nonumber\\
    &\quad\quad\quad +  U^{(F,\eta\eta')}_{mn bb';m'n'b''b'''}(\kk,\kk') O_{b'' m'\eta' s',b''' n'\eta s}(\kk')\Big] \label{eq:HMF}\\
    U^{(H,\eta\eta'')}_{mnbb';m'n'b''b'''}(\kk,\kk')&=\sum_{\GG} U^{(\eta\eta'')}_{mn,m'n'}(\QQ_b-\QQ_{b'};\kk+\QQ_{b'},\kk'+\QQ_{b'''})\delta_{\QQ_b-\QQ_{b'}+\QQ_{b''}-\QQ_{b'''},\GG}\label{eq:HMF-Hartree}\\
    U^{(F,\eta\eta')}_{mn bb';m'n'b''b'''}(\kk,\kk') &=-\sum_{\GG} U^{(\eta\eta')}_{mn',m'n}(\kk-\kk'-\QQ_{b'''}+\QQ_b;\kk'+\QQ_{b'''},\kk+\QQ_{b'})\delta_{\QQ_b-\QQ_{b'}+\QQ_{b''}-\QQ_{b'''},\GG}\label{eq:HMF-Fock}
\end{align}

The interaction energy can be written as 
\begin{align}
    \langle \hH_I\rangle = \frac{1}{2} \frac{1}{\Omega_{tot}}\sum_{\kk\in \mrm{fMBZ}}\sum_{bb'}\sum_{m\eta s,n\eta's'} (H^{MF}_{I})_{b m\eta s,b' n\eta' s'}(\kk) O_{b m\eta s,b' n\eta' s'}(\kk) + const' \label{eq:Eint2}
\end{align}
where we have also neglected a constant term that does not depend on order parameters. Together with $\hH_0$ (\cref{eq:H0-band-basis}) in the fMBZ
\begin{align}
    (H_0)_{bm\eta s,b'm'\eta' s'}&=\epsilon^{(\eta)}_{m}(\kk+\QQ_b)\delta_{bb'}\delta_{mm'}\delta_{\eta\eta'}\delta_{ss'}\, .\label{eq:H0-band-basis-folded}
\end{align}
The matrix element for total mean field Hamiltonian is $(H_0)_{bm\eta s,b'm'\eta' s'}+\Delta_{bm\eta s,b'm'\eta's'}$ and the total energy is 
\begin{align}
    E_{tot} = \mrm{Tr}\left[\left(H_0+\frac{1}{2}H^{MF}_{I}\right)O^T\right]\, 
\end{align}
where the trace is taken over spin, band, momentum in fMBZ, and the translation-symmetry-breaking momentum.

\end{widetext}

\subsection{Classifying the order parameters\label{app:order-sym-prop}} 
In this section we prove that $\mS_V,\mS_{s\pm}$ are invariant under U(2)$\times$U(2) symmetry. The invariance under charge U(1) is trivial.
To see the other symmetry properties, we notice that $||A||^2\equiv\mrm{Tr}(A^\dagger A)$ is invariant when $A$ undergoes unitary transformation $A\to UAU^\dagger$. We will drop the index of $\kk$, $b$ temporarily as these symmetry operations does not change momentum. Thus if $[\Theta,\Theta']=0$ under rotation 
$ \nnorm{[O,\Theta]}^2 \to \nnorm{[e^{i\theta\Theta'}Oe^{-i\theta\Theta'},\Theta]}^2=
    \nnorm{e^{i\theta\Theta'}[O,\Theta]e^{-i\theta\Theta'}}^2
    = \nnorm{[O,\Theta]}^2
$
is also invariant. From this one can easily see $\mS_V$ is not affected by U(1)$_V$, SU(2)$_\pm$, and $\mS_{s\pm}$ are not affected by U(1)$_V$. 

We then consider the transformation of $\mS_{s\pm}$ under spin-SU(2). We first prove the useful identity
\begin{align}
    \sum_i\nnorm{[O,\sigma\tau \varsigma_i]}^2=\sum_i\nnorm{[O,\sigma\tau e^{-i\bm{\theta}\cdot \bm{\varsigma}}\varsigma_ie^{i\bm{\theta}\cdot \bm{\varsigma}}]}^2\label{eq:rot-spin-only}
\end{align}
for any two by two Hermitian matrix $\sigma,\tau$ and $\bm{\theta}=(\theta_1,\theta_2,\theta_3),\bm{\varsigma}=(\varsigma_x,\varsigma_y,\varsigma_z)$. \cref{eq:rot-spin-only} hold for any finite $\bm{\theta}$ if it is valid for infinitesimal $\bm{\theta}$, which can be verified by keeping only the first order term of $\bm{\theta}$ as
\begin{align}
    \sum_i&\nnorm{[O,\sigma\tau e^{-i\bm{\theta}\cdot \bm{\varsigma}}\varsigma_ie^{i\bm{\theta}\cdot \bm{\varsigma}}]}^2 - \sum_i\nnorm{[O,\sigma\tau \varsigma_i]}^2 \nonumber \\
    & = \sum_i\nnorm{[O,\sigma\tau \left(s_i-i[\bm{\theta}\cdot \bm{\varsigma},\varsigma_i]\right)]}^2 - \sum_i\nnorm{[O,\sigma\tau \varsigma_i]}^2 \nonumber\\
    &= \sum_i 2{\rm Tr}\left([O,\sigma\tau \varsigma_i][O,\sigma\tau[\bm{\theta}\cdot \bm{\varsigma},\varsigma_i] ]\right) \nonumber\\
    &= \sum_{ijk} 2\theta_j\epsilon_{jik}{\rm Tr}\left([O,\sigma\tau \varsigma_i][O,\sigma\tau \varsigma_k ]\right)\nonumber\\
    &= 0 \mrm{\;(\;}i\mrm{\leftrightarrow }k\mrm{\;and\;\;Tr}AB=\mrm{Tr}BA)\;.
\end{align}
We then calculate $\mS_{s\pm}$ under spin-SU(2)$_\pm$:
\begin{align}
    \mS' & =\frac{1}{4}\sum_{i}\nnorm{\left[e^{i\sigma_0\tau_{v'} \bm{\theta}\cdot\bm{\varsigma}}Oe^{-i\sigma_0\tau_{v'} \theta\cdot \bm{\varsigma}},\sigma_0\tau_v \varsigma_i\right]}^2 \nonumber \\
    & = \frac{1}{4}\sum_{i}\big|\big|e^{i\sigma_0\tau_{v'} \bm{\theta}\cdot\bm{\varsigma}}\left[O,e^{-i\sigma_0\tau_{v'} \bm{\theta}\cdot\bm{\varsigma}}\sigma_0\tau_v \varsigma_ie^{i\sigma_0\tau_{v'} \bm{\theta}\cdot\bm{\varsigma}}\right]\nonumber\\
    &\qquad\qquad\qquad\qquad e^{-i\sigma_0\tau_{v'} \bm{\theta}\cdot\bm{\varsigma}}\big|\big|^2 \nonumber \\
    & = \frac{1}{4}\sum_{i}\nnorm{\left[O,\sigma_0e^{-i\sigma_0\tau_{v'} \bm{\theta}\cdot\bm{\varsigma}}\tau_v \varsigma_ie^{i\sigma_0\tau_{v'} \bm{\theta}\cdot\bm{\varsigma}}\right]}^2 \label{eq:su2-transform}
\end{align}
where $v,v'=\pm$. Since $\tau_{v'}^2=\tau_{v'}$ for $v'=\pm$, one has $e^{i\tau_{v'}\bm{\theta}\cdot\bm{\varsigma} } = \sum_{n=0}^\infty\frac{(i\tau_{v'}\bm{\theta}\cdot\bm{\varsigma})^n}{n!}=\tau_0\varsigma_0+\tau_{v'}\sum_{n=1}^\infty\frac{(i\bm{\theta}\cdot\bm{\varsigma})^n}{n!}=\tau_0\varsigma_0-\tau_{v'}\varsigma_0+\tau_{v'}e^{i\bm{\theta}\cdot\bm{\varsigma}}$.Then we have 
\begin{align}
    e^{-i\tau_{v'}\bm{\theta}\cdot\bm{\varsigma}}\tau_v\varsigma_i e^{i\tau_{v'}\bm{\theta}\cdot\bm{\varsigma}}=\tau_v\left[\left(\tau_0-\tau_{v'}\right)\varsigma_i+\tau_{v'}e^{-i\bm{\theta}\cdot\bm{\varsigma}}\varsigma_ie^{i\bm{\theta}\cdot\bm{\varsigma}}\right]\label{eq:valley-spin}
\end{align}
The physical meaning is simple: $\tau_{v'}$ is a projection matrix, and we only perform spin-SU(2) rotation on the subspace that $\tau_{v'}$ projecting to, and the action on the complement is identity. Then making use of \cref{eq:rot-spin-only,eq:su2-transform,eq:valley-spin} we have
\begin{itemize}
    \item for $v=v'=\pm$, which corresponds to $\mS_{s\pm}$ transformed under spin SU(2) rotation in $\pm$ valley,$\mS_{s\pm}'=\sum_i\nnorm{[O,\sigma_0\tau_\pm e^{-i\theta\cdot s}s_ie^{i\theta\cdot s}]}^2=\sum_i\nnorm{[O,\sigma_0\tau_\pm s_i]}^2=\mS_{s\pm}$
    \item for $v=-v'=\pm$, which corresponds to $\mS_{s\pm}$ transformed under spin SU(2) rotation in $\mp$ valley $\mS_{s\pm}'=\sum_i\nnorm{[O,\sigma_0\tau_\pm s_i]}^2=\mS_{s\pm}$
\end{itemize}

Performing a gauge transformation among the two flat bands in valley $v$ means a unitary transformation of the order parameter $O\to e^{-i\bm{\theta}\cdot\bm{\sigma}\tau_v}Oe^{i\bm{\theta}\cdot\bm{\sigma}\tau_v}$ when $v=\pm$. We can prove that $\mS_{V,s\pm}$ are also invariant under this gauge transformation. The matrix $\Theta$ for generators included in $\mS_{V,s\pm}$ takes the form $\sigma_0\tau_{v}\varsigma_{\mu}$ for $v=0,\pm$ and $\mu=0,x,y,z$. We have
\begin{align}
    &\nnorm{[e^{-i\bm{\theta}\cdot\bm{\sigma}\tau_{v}}Oe^{i\bm{\theta}\cdot\bm{\sigma}\tau_{v}},\sigma_0\tau_{v'}\varsigma_\mu]}^2 \nonumber \\
    &\qquad = \nnorm{[O,\sigma_0e^{i\bm{\theta}\cdot\bm{\sigma}\tau_{v}}\tau_{v'}e^{-i\bm{\theta}\cdot\bm{\sigma}\tau_{v}}\varsigma_\mu]}^2\nonumber\\
    &\qquad = \nnorm{[O,\sigma_0\tau_{v'}\varsigma_\mu]}^2
\end{align}
we have used $[\tau_v,\tau_{v'}]=0$ in the second line.

\subsection{Degeneracy from the HF energy functional\label{app:degenerate}}
There are some different states that are not related by symmetries but are degenerate at the HF level, which has been mentioned in earlier results in TBG \cite{zhang_correlated_2020,kwan_strong-coupling_2023}. For example, the authors of Ref.\cite{zhang_correlated_2020} had found that at $\nu=0$ there are valley-polarized states, spin-polarized states and spin-valley-locked states which are degenerate and they are named the generalized ferromagnetic insulating states (FMI). Here, we prove that for state with order parameters $O_{bm\eta s,b'm'\eta' s}(\kk)$ diagonalized in spin degree of freedom, applying spinless $T_0$ on one spin sector does not change the total energy. It is worth noticing that this operation is not a symmetry as the Wigner theorem \cite{wigner_normal_2004} states that any symmetry should be either a unitary operator or an anti-unitary operator, which is not satisfied by this operation.  Furthermore, by SU(2)$_+\times$SU(2)$_-$ symmetry for all the spin-polarized states that are not polarized in $z$ direction, we can also rotate the spin polarization to $z$ direction and apply the same operation. Thus, two sets of states are related.

Under this operation we obtain a new order paraters $O'$ where $O'_{bm\eta \uparrow,b'm'\eta' \uparrow}(\kk)=O_{bm\eta \uparrow,b'm'\eta' \uparrow}(\kk) $ and $O'_{bm\eta \downarrow,b'm'\eta' \downarrow}(\kk) = O^*_{\bar{b}m\eta \downarrow,\bar{b}'m'\eta' \downarrow}(-\kk) $. We have use the notation $\bar{b}$ such that $\QQ_{\bar{b}}=-\QQ_{b}$ module reciprocal lattice. We write the order parameters as $O=O_{\uparrow\uparrow}\delta_{s,\uparrow}+O_{\downarrow\downarrow}\delta_{s,\downarrow}$. We demonstrate the three terms in total energy are unchanged one by one:
\begin{itemize}
    \item the kinetic term. $\hH_0$ is diagonalized in spin subspace, and the total kinetic energy is the summation of the kinetic energy of spin up and spin down subspace, and each of them is unchanged under $T_0$.
    \item the Hartree term. For the Hartree term, this proof consists of two steps: 1. $O$ and $O'$ give the same Hartree potential, which preserves $T_0$ and is diagonal in the spin subspace. 2. For Hamiltonian preserving $T_0$ and diagonal in spin subspace, we repeat the argument for the kinetic term.
    \item the Fock term. From \cref{eq:HMF-Fock} one can see that when the order parameters are diagonalized in spin subspace, the Fock term of mean field Hamiltonian are also diagonalized in spin subspace, by which we write it as $\Delta^F_{\uparrow\uparrow}$ and $\Delta^F_{\downarrow\downarrow}$ and $\Delta^F_{ss}$ depends only on $O_{ss}$. As $T_0$ is the symmetry of the Hamiltonian, acting $T_0$ does not change the Fock energy $ \frac{1}{2}\sum_s {\rm Tr}[\Delta^F_{ss}O_{ss}] $. Meanwhile $T_0$ does not flip spin, so each term $ \frac{1}{2} {\rm Tr}[\Delta^F_{ss}O_{ss}] $ is seperately invariant when we apply $T_0$ to each spin sector.

\end{itemize}

\end{document}